\documentclass[12pt]{article}
\usepackage{amsmath,enumerate,amsfonts,amssymb,amsthm,enumerate,graphicx,url,color,ulem}
\def \tblb{\\ \hline}

\definecolor{black}{rgb}{0,0,0}
\newcommand{\bl}[1]{\textcolor{black}{#1}}
\usepackage{graphicx}
\usepackage{amssymb,amsmath,amsfonts}
\usepackage{epsf}
\usepackage{wrapfig}
\newtheorem{algorithm}{Algorithm}

\newcommand{\R}{\mathbb{R}}
\newcommand{\Z}{\mathbb{Z}}

\newcommand{\E}{\mathbb{E}}

\title{
Stochastic Representations of  Ion Channel  Kinetics and Exact Stochastic Simulation of Neuronal Dynamics}
\author{David F. Anderson$^{1}$, Bard Ermentrout$^{2}$, and Peter J. Thomas$^{3}$\\
$^1$University of Wisconsin, Department of Mathematics\\
$^2$University of Pittsburgh, Department of Mathematics\\
$^3$Case Western Reserve University, Department of Mathematics, \\ Applied Mathematics, and Statistics}

\begin{document}
\maketitle
\abstract{
In this paper we provide two representations for stochastic ion channel kinetics, and compare the performance of exact simulation with \bl{a commonly used numerical approximation strategy}.   The first representation we present is a random time change representation, popularized by Thomas Kurtz, with the second being analogous to a ``Gillespie'' representation.   Exact stochastic algorithms are provided for the different representations, which are preferable to either (a)  fixed time step or (b) piecewise constant propensity algorithms, which still appear in the literature.  As examples, we provide versions of the exact algorithms for the Morris-Lecar conductance based model, and detail the error induced, both in a weak and a strong sense, by the use of approximate algorithms on this model.  We include ready-to-use implementations of the random time change algorithm in both XPP and Matlab. Finally, through the consideration of parametric sensitivity analysis, we show how the representations presented here are useful in the development of further computational methods.   The general representations and simulation strategies provided here are  known in other parts of the sciences, but less so in the present setting.

}

%{New revisions as of Jan.~10 in blue -- Peter}\\
%\dfa{Dave's suggestions on January 16-19th in red (selected)}\\
%{January 29 revisions -- Peter}\\
%\dfab{Feb. 3rd revisions -- Dave}

\section{Introduction}

Fluctuations in membrane potential arise in part due to stochastic switching in voltage-gated ion channel populations \cite{Dorval+White:2005:JNsci,LaingLord2010,White+Kay+Rubinstein:2000:TrendsNsci}.  We consider a stochastic modeling, i.e. master equation, framework \cite{AndersonKurtz2010Chapter,Colquhoun+Hawkes:1983chapter,EarnshawKeener2010SIADS,LeeOthmer2010JMB,Wilkinson2011}  for neural dynamics, with noise arising through the molecular fluctuations of ion channel states.  
We consider model nerve cells that may be represented by a single isopotential volume surrounded by a membrane with capacitance $C>0$.  Mathematically, these are hybrid stochastic models which include components, for example the voltage, that are continuous and piecewise differentiable and components, for example  the number of open  potassium channels, that make discrete transitions or jumps. \bl{These components are coupled because the parameters of the ODE for the voltage depend upon the number of open  channels, and the propensity for the opening and closing of the channels depends explicitly upon the time-varying voltage.}

These \bl{hybrid stochastic} models are typically described in the neuroscience literature by providing an ODE governing the absolutely continuous portion of the system, which is valid between jumps of the discrete components, and a chemical master equation providing the dynamics of the probability distribution of the jump portion, which itself depends upon the solution to the ODE.   \bl{These models are piecewise-deterministic Markov processes (PDMP) and one can therefore also characterize them by providing (i) the ODE for the absolutely continuous portion of the system and (ii) both a rate function that determines when the next jump of the process occurs and a transition measure determining which type of jump occurs at that time \cite{Davis:1984}.   Recent works making use of the PDMP formalism has led to limit theorems \cite{PTW:2010,PTW:2012}, dimension reduction schemes \cite{WTP:2012}, and extensions of the models to the spatial domain \cite{BR:2011,RiedlerThieullenWainrib2012ElecJProb}.}

 In this paper, we take a different perspective.  
We will introduce here two pathwise stochastic representations for these models that are similar to It\^o SDEs or Langevin models.  The difference between the models presented here and Langevin models is that here
% with the difference being that 
the noise arises via stochastic counting processes as opposed to Brownian motions.
These representations give a different type of insight into the models than master equation representations do, and, in particular, they imply different exact simulation strategies.  These strategies are well known in some parts of the sciences, but less well known in the current context  \cite{AndersonKurtz2010Chapter}.\footnote{For an early statement of an exact algorithm for the hybrid case in a neuroscience context see  (\cite{ClayDeFelice1983BiophysJ}, Equations 2-3).  Strassberg and DeFelice  further investigated circumstances under which it is possible for random microscopic events (single ion channel state transitions) to generate random macroscopic events (action potentials) \cite{StrassbergDeFelice1993NeCo} using an exact simulation algorithm.  \bl{Bressloff,} Keener, and Newby  used an exact algorithm in a recent study of channel noise dependent action potential generation in the Morris-Lecar model \cite{KeenerNewby2011PRE,NewbyBressloffKeener2013PRL}. \bl{For a detailed introduction to stochastic ion channel models, see \cite{GroffDeRemigioSmith2009Chapter,SmithKeizer2002BookChapter}.}} 

 {From a computational standpoint the change in perspective from the master equation to pathwise  representations is useful for a number of reasons.  First, the different representations naturally imply different exact simulation strategies.  Second, and perhaps more importantly, the different representations themselves can be utilized to develop new, highly efficient, computational methods such as finite difference methods for the approximation of parametric sensitivities, and multi-level Monte Carlo methods for the approximation of expectations \cite{AndCFD2012,AndHigham2012,AHS2014}.  Third, the representations can be utilized for the rigorous analysis of different computational strategies and for the algorithmic reduction of models with multiple scales \cite{AndersonGangulyKurtz2011AnnApplPr,AK2014,Ball06,KurtzKang:2013}.}
 
% they can be utilized since the standard Gillespie algorithm, in which propensity functions are constant between discrete events, does not provide an exact simulation strategy when the propensities of the jump process depend upon the solution of a nontrivial ODE, as in the current case.}

%From a computational standpoint this change in perspective is useful since the standard Gillespie algorithm, in which propensity functions are constant between discrete events, does not provide an exact simulation strategy when the propensities of the jump process depend upon the solution of a nontrivial ODE, as in the current case.

We note that the types of representations and simulation strategies highlighted here are  known in other branches of the sciences, especially operations research and queueing theory \cite{Glynn89,Haas2002}, and stochastically modeled biochemical processes \cite{Anderson2007JChemPhys,AndersonKurtz2010Chapter}.  See also \cite{Riedler2013} for a  treatment of such stochastic hybrid systems and some corresponding approximation algorithms. However, {the representations} are not as well known in the context of computational neuroscience and as a result approximate methods {for the simulation of sample paths} including (a) fixed time step methods, or (b) piecewise constant propensity algorithms, are still utilized in the literature in situations where there is no need to make such approximations. 
%\footnote{{For recent examples see \cite{SchmandtGalan2012PRE} (fixed time step binomial approximation) and  \cite{FischSchwalgerLindnerHerzBenda2012JNSci,KisperskyTilman2008Scholarpedia,SchwalgerFischBendaLindner2010PLosCB}  (piecewise constant propensity approximation).} {I am uncomfortable pointing to specific papers and saying they are doing something wrong.  I would say we cut this footnote, and simply leave it in the main article text that there are still approximate algorithms appearing in the literature.}}  
Thus, the main contributions of this paper are: $(i)$ the formulation of the two pathwise representations for the specific models under consideration,  $(ii)$ {a presentation of the corresponding exact simulation strategies for the different representations}, and $(iii)$ a comparison of the error induced by utilizing an approximate simulation strategy on the Morris-Lecar model.  {Moreover, we show how to utilize the different representations in the development of  methods for parametric sensitivity analysis}.   
%The representations, on top of providing new insight into the models themselves, can be utilized in future work in multi-scale analysis \cite{Ball06},  the rigorous development of approximate simulation strategies \cite{AndersonGangulyKurtz2011AnnApplPr}, and new multi-level simulation strategies \cite{AndHigham2012}.

The outline of the paper is as follows.  In Section, \ref{sec:development} we heuristically  develop two distinct  representations and provide the corresponding numerical methods.  In Section \ref{sec:ML}, we present the random time change representation as introduced in Section \ref{sec:development} for a particular conductance based model,  the planar Morris-Lecar model, with a single ion channel treated stochastically.  Here, we illustrate the corresponding numerical strategy on this example and provide in the appendix both XPP  and Matlab code for its implementation.  In Section \ref{sec:general}, we present an example of a conductance based model with more than one stochastic gating variable, namely the Morris-Lecar model with both ion channel types (calcium channels and potassium channels) treated stochastically.
%{As an example of general conductance based models such as the  Hodgkin-Huxley system, in Section \ref{sec:general} we consider the Morris-Lecar model with two ion channel types treated stochastically (DFA: Please check this last sentence to make sure it is saying what you wanted)}. 
To illustrate both the strong and weak divergence between the exact and approximate algorithms, in Section \ref{sec:convergence} we compare trajectories and histograms generated by the exact algorithms and the piecewise constant approximate algorithm. In Section \ref{sec:coupling}, we  show how to utilize the different representations presented here in the development of methods for parametric sensitivity analysis, which is a powerful tool for determining parameters to which a system output is most responsive.
In Section \ref{sec:future}, we provide conclusions and discuss avenues for future research.

%
%\vspace{.1in}
%
%\hrule
%
%\vspace{.1in}
%
%\noindent \textbf{These are just notes}
%\begin{enumerate}
%	\item DAVE In section 2, we develop the representation for a simple model
%	\[
%		A \overset{\alpha}{\underset{\beta}{\rightleftarrows}} B
%	\]
%	Next, give development with $\alpha(t), \ \beta(t)$.
%	\[
%	A \overset{\alpha(t)}{\underset{\beta(t)}{\rightleftarrows}} B
%	\]
%	\item PETER In Section 3, we give explicit equations for Morris-Lecar.  We illustrate the numerical strategy on this example and appendix provides both XPP  and Matlab code for this model.
%	\item PETER In Section 4, we will give a more general representation. Start with HH.
%	\item DAVE In Section 5 , we discuss future ideas:  this opens up a whole world of computational possibilities including tau-leaping, efficiency issues, etc.  Here we show an algorithm, which is exact, and therefore has intrinsic value.  The other methods are, of course, good but each has a bias which is difficult to quantify precisely.  
%\end{enumerate}
%In general, a cell membrane contains a variety of ion channels of a finite number of different types.   
%
%
%\vspace{.1in}
%
%\hrule
%
%\vspace{.1in}

\section{Two stochastic representations}
\label{sec:development}

The purpose of this section is to heuristically present two pathwise representations for the relevant models.  In Section \ref{sec:RTC} we present the random time change representation.  In Section \ref{sec:Gillespie} we present a ``Gillespie'' representation, which is analogous to the PDMP formalism discussed in the introduction.  In each subsection, we provide the corresponding numerical simulation strategy.  See  \cite{AndersonKurtz2010Chapter} for a development of the representations in the biochemical setting and see  \cite{EK:1986,Kurtz80,KurtzPop81} for a rigorous mathematical development.

\subsection{Random time change representations}
\label{sec:RTC}

%{Note: Customarily, $\alpha$ denotes a channel opening rate and $\beta$ a closing rate, so in the example below it is more natural to call $A$ the ``closed" state and $B$ the ``open" state.  I revised this section accordingly.}

%In this section we develop the random time change representation for our models of interest.  For a more complete overview of this representation, see \cite{AndersonKurtz2010Chapter}.  

We begin with an example.  Consider a model of a system that can be in one of two states, $A$ or $B$, which  represent a ``closed'' and an ``open'' ion channel, respectively.  We model the dynamics of the system by assuming that the dwell times in states $A$ and $B$ are determined by independent  exponential random variables with  parameters $\alpha>0$ and $\beta>0$, respectively.  A graphical representation for this model is 
\begin{equation}\label{eq:simple_model}
	A \overset{\alpha}{\underset{\beta}{\rightleftarrows}} B.
\end{equation}
The usual formulation of the stochastic version of this model proceeds by assuming that the probability that a closed channel opens in the next increment of time $\Delta s$ is  $\alpha \Delta s + o(\Delta s)$, whereas the probability that an open channel closes is $\beta \Delta s + o(\Delta s)$.  This type of stochastic model is often described mathematically by providing the ``chemical master equation,'' which  for \eqref{eq:simple_model} is simply
\begin{align*}
	\frac{d}{dt} p_{x_0}(A,t) &= -\alpha p_{x_0}(A,t) + \beta p_{x_0}(B,t)\\
	\frac{d}{dt} p_{x_0}(B,t) &=-\beta p_{x_0}(B,t) + \alpha p_{x_0}(A,t),
\end{align*}
where $p_{x_0}(x,t)$ is the probability of being in state $x\in \{A,B\}$ at time $t$ given an initial condition of state $x_0$.  Note that the chemical master equation is a linear ODE governing the dynamical behavior of the \textit{probability distribution} of   the model, and does not provide a stochastic representation for a particular \textit{realization} of the process.

In order to construct such a pathwise representation, let $R_1(t)$ be the number of times the transition $A\to B$ has taken place by time $t$ and, similarly, let $R_2(t)$ be the number of times the transition $B\to A$ has taken place by time $t$.  We let $X_1(t)\in \{0,1\}$ be one  if the channel is closed at time $t$, and zero otherwise, and let $X_2(t) = 1-X_1(t)$ take the value one if and only if the channel is open at time $t$.  Then, denoting $X(t) = (X_1(t),X_2(t))^T$, we  have
\begin{equation*}
	X(t) = X(0) + R_1(t)\left(\begin{array}{c}
		-1\\
		1
	\end{array}\right)  + R_2(t)\left(\begin{array}{c}
		1\\
		-1
	\end{array}\right).
\end{equation*}

We now consider how to represent the counting processes $R_1,R_2$ in a useful fashion, and we do so with unit-rate Poisson processes as our mathematical building blocks.  We recall that a unit-rate Poisson process can be  constructed in the following manner \cite{AndersonKurtz2010Chapter}.  Let $\{e_i\}_{i=1}^\infty$ be independent exponential random variables with a parameter of one.  Then, let $\tau_1  = e_1, \tau_2 = \tau_1 + e_2, \cdots, \tau_n = \tau_{n-1} + e_n$, etc.  The associated unit-rate Poisson process, $Y(s)$, is simply the counting process determined by the number of points $\{\tau_i\}_{i=1}^\infty$, that come before $s\ge 0$.  For example, if we let ``x'' denote the points $\tau_n$ in the image below
\begin{center}\
  \begin{tabular}{|l|l} \hspace{0.5in}x \hspace{0.1in}x
    \hspace{0.8in}x \hspace{0.4in}x \hspace{0.2in}x \hspace{0.3in}x&
    \hspace{0.1in} \hspace{0.2in}x \hspace{0.4in}x 
    \hspace{0.3in}
    \tblb
    &$s$ \\
  \end{tabular}
\end{center}
then $Y(s) = 6$. 

\bl{Let $\lambda: [0,\infty) \to \R_{\ge 0}$.}  If instead of moving at constant rate, $s$, along the horizontal axis, we move instead at rate $\lambda(s)$, then the number of points observed by time $s$ is  $Y\left(\int_0^s \lambda(r) dr\right)$.  Further, from basic properties of exponential random variables, whenever $\lambda(s) >0$ the probability of seeing a jump within the next small increment of time $\Delta s$ is
\[
	P\left( Y\left( \int_0^{s + \Delta s} \lambda(r) dr\right) - Y\left( \int_0^{s} \lambda(r) dr\right)\ge 1\right) \approx \lambda(s)\Delta s.
\]
%where $\F_s$ should be thought of as the ``information'' we have by observing the process up through time $s$.\footnote{Technically, $\F_s$ is a filtration, i.e. an increasing sequence of $\sigma$-algebras.}
% is the filtration representing all of the information contained in the Poisson process $Y$ up through time $\int_0^s \lambda(r) dr.$\footnote{Mathematically, $\F_s = \mathcal G_{\int_0^s \lambda(r) dr}$, where $\mathcal G_t$ is the filtration generated by $Y$.}  
Thus, the \textit{propensity} for seeing another jump is precisely $\lambda(s)$.  

Returning to the discussion directly following \eqref{eq:simple_model}, and noting that $X_1(s) + X_2(s) = 1$ for all time, we note that the propensities of reactions 1 and 2 are
\[
	\lambda_1(X(s)) = \alpha X_1(s), \quad \lambda_2(X(s)) = \beta X_2(s).
\]
%where $1\{S\}$ is the indicator function for the event $S$. %\footnote{PJT: I added this explanation.  Is it OK?  Do we need to mention, in another footnote, that ``events" are sets that are measurable with respect to some sigma algebra?  Is it the Borel $\sigma$-algebra on the real line, raised to $M^{\text{th}}$ power, where $M$ is the number of reactions, i.e.~the number of Poisson processes?}.
Combining all of the above implies that we can represent $R_1$ and $R_2$ via 
\[
	R_1(t) = Y_1\left( \int_0^t \alpha X_1(s) \, ds\right), \quad R_2(t) = Y_2\left( \int_0^t \beta X_2(s) \, ds\right),
\]
and so a pathwise representation for the stochastic model \eqref{eq:simple_model} is
\begin{equation}\label{eq:simple_model_pathwise}
	X(t) = X_0 + Y_1\left( \int_0^t \alpha X_1(s) \, ds\right)\left(\begin{array}{c}
		-1\\
		1
	\end{array}\right) + Y_2\left( \int_0^t \beta X_2(s) \, ds\right)\left(\begin{array}{c}
		1\\
		-1
	\end{array}\right),
\end{equation}
where $Y_1$ and $Y_2$ are independent, unit-rate Poisson processes.

Suppose now that $X_1(0) + X_2(0) = N \ge 1$.  For example, perhaps we are now modeling the \textit{number} of open and closed ion channels out of a total of $N$, as opposed to simply considering a single such channel.  Suppose further that the propensity, or rate, at which ion channels are opening can be modeled as
\[
	\lambda_1(t,X(t)) = \alpha(t) X_1(t)
\] 
and the rate at which they are closing has propensity 
\[
	\lambda_2(t,X(t)) = \beta(t)X_2(t),
\]
where $\alpha(t),\beta(t)$ are non-negative functions of time, perhaps being voltage dependent.  {That is, suppose that for each $i \in \{1,2\}$, the conditional probability of seeing the counting process $R_i$ increase in the interval $[t,t+h)$ is $\lambda_i(t,X(t))h + o(h)$.}
%determining the number of times the $i$th reaction has taken place 
%\begin{align*}
%	P(R_i(t+h) - R_i(t) = 0 \ | \ \F_t^X) &= 1-\lambda_i(t,X(t)) h + o(h)\\
%	P(R_i(t+h) - R_i(t) = 1 \ | \ \F_t^X) &= \lambda_i(t,X(t)) h + o(h),
%\end{align*}
%and
%\[
%	P(R_1(t+h) - R_1(t) >0, R_2(t+h) - R_2(t) >0  \ | \ \F_t^X) = o(h),
%\]
%where $\F_t^X$ represents the information we have by observing the process $X$ up to time $t$.\footnote{Technically, $\F_t^X$ is a filtration, i.e. an increasing sequence of $\sigma$-algebras.  Readers unfamiliar with this terminology can simply think of $\F_t^X$ as ``information.''}  
%{Then, analogously with before,} 
{The expression analogous to \eqref{eq:simple_model_pathwise} is now}%Under these assumptions,  we have \cite{AndersonKurtz2010Chapter}
\begin{equation}
	X(t) = X_0 + Y_1\left( \int_0^t \alpha(s) X_1(s) \, ds\right)\left(\begin{array}{c}
		-1\\
		1
	\end{array}\right) + Y_2\left( \int_0^t \beta(s) X_2(s) \, ds\right)\left(\begin{array}{c}
		1\\
		-1
	\end{array}\right).
\end{equation}

Having motivated the time dependent representation with the simple model above, we turn to the more general context.  We now assume a jump model consisting of $d$ chemical constituents (or ion channel states) undergoing transitions determined via $M>0$ different reaction channels.  For example, in the toy model above, the chemical constituents were $\{A,B\}$, and so $d = 2$, and the reactions were $A \to B$ and $B\to A$, giving $M = 2$.  We suppose that $X_i(t)$ determines the value of the $i$th constituent at time $t$, so that $X(t) \in \Z^d$, and that the propensity  function of the $k$th reaction is $\lambda_k(t,X(t))$.  We further suppose that if the $k$th reaction channel takes place at time $t$, then the system is updated according to addition of the reaction vector $\zeta_k \in \Z^d$,
 \[
	X(t) = X(t-) + \zeta_k.
\]
The associated pathwise stochastic representation for this model is
\begin{equation}\label{eq:RTC_main}
	X(t) = X_0 + \sum_k Y_k\left(\int_0^t \lambda_k(s,X(s)) \, ds\right)\zeta_k,
\end{equation}
where the $Y_k$ are independent unit-rate Poisson processes.  The chemical master equation for this general model is
\begin{align*}
	\frac{d}{dt}P_{X_0}(x,t) &= \sum_{k=1}^M P_{X_0}(x - \zeta_k,t)\lambda_k(t,x-\zeta_k) - P_{X_0}(x,t)\sum_{k=1}^M \lambda_k(t,x),
\end{align*}
where $P_{X_0}(x,t)$ is the probability of being in state $x\in \Z^d_{\ge 0}$ at time $t\ge 0$ given an initial condition of $X_0.$

When the  variable $X\in\Z^d$ represents the randomly fluctuating state of an ion channel in a single compartment conductance based neuronal model, we include the membrane potential $V\in\R$ as an additional dynamical variable.  In contrast with neuronal models incorporating Gaussian noise processes, here we consider the voltage to evolve deterministically, conditional on {the states of one or more ion channels.  For illustration, suppose we have a single ion channel type with state variable $X$.  Then,} we supplement the pathwise representation (\ref{eq:RTC_main}) with the solution of a differential equation obtained from Kirchoff's current conservation law:
\begin{align}\label{eq:voltage}
C\frac{dV}{dt}=I_{\text{app}}(t)-I_V(V(t))-\left(\sum_{i=1}^d g^o_i X_i(t)\right)(V(t)-V_X)
\end{align}
Here, $g^o_i$ is the conductance of an individual channel when it is the $i^{\text{th}}$ state, for $1\le i \le d$. The sum gives the total conductance associated with the channel represented by the vector $X$; the reversal potential for this channel is the constant $V_X$.  The term $I_V(V)$ captures {any deterministic} voltage-dependent currents due to other channels besides channel type $X$, and $I_{\text{app}}$ represents a time-varying, deterministic applied current.    In this case the propensity function will explicitly be a function of the voltage and we may replace $\lambda_k(s,X(s))$ in \eqref{eq:RTC_main} with $\lambda_k(V(s),X(s))$.  {If multiple ion channel types are included in the model then, provided there are a finite number of types each with a finite number of individual channels, the vector $X\in\Z^d$  represents the aggregated channel state.   For specific examples of handling a single or multiple ion channel types, see Sections \ref{sec:ML} and \ref{sec:general}, respectively.}

\subsubsection{Simulation of the representation \eqref{eq:RTC_main}-\eqref{eq:voltage}}
\label{sec:sim_RTC}

The construction of the representation \eqref{eq:RTC_main} provided above implies a simulation strategy in which each point of the Poisson processes $Y_k$, denoted $\tau_n$ above, is generated sequentially and as needed.  The time until the next reaction that occurs past time $T$ is simply 
\[
	\Delta =\min_k\left\{ \Delta_k :  \int_0^{T + \Delta_k} \lambda_k(s,X(s)) \, ds = \tau_{T}^k\right\},
\]
where $\tau_{T}^k$ is the first point associated with $Y_k$ coming after $\int_0^T \lambda_k(s,X(s)) \, ds$:
\[
	\tau_T^k \equiv \inf\left\{r > \int_0^T \lambda_k(s,X(s)) \, ds : Y_k(r) - Y_k\left(\int_0^T \lambda_k(s,X(s)) \, ds\right) = 1\right\}.
\]
  The reaction that took place is indicated by the index at which the minimum is achieved.   See \cite{Anderson2007JChemPhys} for more discussion on this topic, including the algorithm provided below in which $T_k$ will denote the {value of the} integrated intensity function 
% this could be called the "internal time"?
$\int_0^t \lambda_k(s,X(s))\, ds$ and $\tau_k$ will denote the first point associated with $Y_k$ located after $T_k$.

\vspace{.125in}
All random numbers generated in the algorithm below are assumed to be independent.

\vspace{.125in}
\begin{algorithm}[{For the simulation of the representation \eqref{eq:RTC_main}-\eqref{eq:voltage}}] $ $
\label{main-algorithm}
  \begin{enumerate}
  \item Initialize: %\footnote{Have to initialize the voltage also.  For any voltage there will be an equilibrium distribution of the channel states, and one could draw the initial channel state from this distribution if desired.}  
  set the initial number of molecules of each species, $X$. Set the initial voltage value $V$.  Set $t = 0$.  For each $k$, set $\tau_k = 0$ and $T_k = 0$.
  \item Generate $M$ independent, uniform(0,1) random numbers $\{r_k\}_{k=1}^M$.  For each $k \in \{1,\dots,M\}$ set $\tau_k = \ln(1/r_k)$.
  \item \bl{Numerically integrate \eqref{eq:voltage} forward in time until one of the following equalities hold:%For each $k \in \{1,\dots,M\}$ set $\Delta_k$ to be the solution \footnote{You have to integrate Equation \ref{eq:voltage} at the same time, so you have the time dependence of the $\lambda_k$.}
  \begin{align}\label{eq:num_int}
  	\int_t^{t+\Delta}\lambda_k(V(s),X(s)){\,ds} = \tau_k - T_k.
\end{align}
    \label{impstep2}
    }
  \item Let $\mu$ be the index of the reaction channel where the equality \eqref{eq:num_int} holds.
  \item {For each $k$, set 
  \[
  	T_k = T_k + \int_t^{t+\Delta} \lambda_k(V(s),X(s))\, ds,
\]}
\bl{where $\Delta$ is determined in Step \ref{impstep2}.}
    \item {Set $t = t + \Delta$ and 
    \[
    	X \leftarrow X +\zeta_\mu. 
%       	X = X +\zeta_\mu. 
    \]}
  \item Let $r$ be uniform(0,1) and set $\tau_{\mu} =
    \tau_{\mu} + \ln(1/r)$.
    \label{modstep2}
  \item Return to step 3 or quit.
  \end{enumerate}
  \end{algorithm}

%\vspace{.1in}
%{Figure \ref{fig:flowchart} illustrates the algorithm graphically.}
%{Conceptually, we may distinguish two classes of time variables. The ``external'' or ``objective'' time $t$, and corresponding intervals $\Delta$, represent times as they would be measured on the laboratory clock.  In contrast, the times $T_k$ and $\tau_k$, that are ``internal'' to the $k^{\text{th}}$ reaction,  measure the passage of time as seen by the $k^{\text{th}}$ Poisson process.}  %{(DFA: I find the flowcharts difficult to parse, and I am not sure how necessary they are.  To me, the algorithm itself is very clear.  I vote we cut the flowcharts.)}

 \bl{Note that with a probability of one,  the index determined in step 4 of the above algorithm is unique at each step.} 
Note also that the above algorithm relies on us being able to calculate a hitting time for each of  the $T_k(t) = \int_0^{t} \lambda_k(s,X(s))\, ds$ exactly.  Of course, in general this is not possible.  However, making use of any reliable integration software  will almost always be sufficient.  \bl{If the equations for the voltage and/or the intensity functions can be analytically solved for, as can happen in the Morris-Lecar model detailed below, then such numerical integration is unnecessary and efficiencies can be gained.}

%\begin{figure}[htbp] %  figure placement: here, top, bottom, or page
%   \centering
%\includegraphics[height=7in,angle=90]{NextReactionAlgorithmFlowchart-v3.pdf}
%%   \includegraphics[width=6in]{flowchart.pdf} 
%   \caption{{Flowchart illustrating the steps of the algorithm for simulation of \eqref{eq:RTC_main}-\eqref{eq:voltage}, the random time change algorithm (Algorithm \ref{main-algorithm}).  The thick line highlights the functional role of the global, ``external'' time variable $t$, in contrast with the time variables $T_k$ and $\tau_k$ that are ``internal'' to the $k^{\text{th}}$ Poisson process. }}
%   \label{fig:flowchart}
%\end{figure}

\subsection{Gillespie representation}
\label{sec:Gillespie}

There are multiple alternative representations  for the general stochastic process constructed in the previous section, with a ``Gillespie'' representation probably being the most useful in the current context.  Following \cite{AndersonKurtz2010Chapter}, we let $Y$ be a unit rate Poisson process and let $\{\xi_i,i = 0,1,2\dots\}$ be independent, uniform $(0,1)$ random variables that are independent of $Y$.  Set 
\[
	\lambda_0(V(s),X(s)) \equiv \sum_{k=1}^M \lambda_k(V(s),X(s)),
\]
$q_0 = 0$ and for $k \in \{1,\dots,M\}$ 
\[
	q_k(s) = \lambda_0(V(s),X(s))^{-1}\sum_{\ell = 1}^k  \lambda_\ell(V(s),X(s)),
\]
where $X$ and $V$ satisfy
\begin{align}
		R_0(t) &= Y\left( \int_0^t \lambda_0(V(s),X(s))\, ds\right)\label{eq:Gill_when} \\ 
		X(t) &= X(0) + \sum_{k=1}^M \zeta_k \int_0^t 1\left\{\xi_{R_0(s-)} \in [q_{k -1}(s-),q_{k}(s-)) \right\} dR_0(s)\label{eq:Gill_where} \\
		C\frac{dV}{dt}&=I_{\text{app}}(t)-I_V(V(t))-\left(\sum_{i=1}^d g^o_i X_i(t)\right)(V(t)-V_X).
		\label{eq:Voltage2}
\end{align}
Then the stochastic process $(X,V)$ defined via \eqref{eq:Gill_when}-\eqref{eq:Voltage2} is a Markov process that is equivalent to \eqref{eq:RTC_main}-\eqref{eq:voltage}, see \cite{AndersonKurtz2010Chapter}.  An intuitive way to understand the above is by noting that $R_0(t)$ simply determines the  holding time in each state, whereas \eqref{eq:Gill_where} simulates the embedded \textit{discrete time} Markov chain (sometimes referred to as the \textit{skeletal chain}) in the usual manner.  Thus, this is the representation for the ``Gillespie algorithm'' {\cite{Gillespie1977}} with time dependent propensity functions \cite{Anderson2007JChemPhys}.
\bl{Note also that this representation is analogous to the PDMP formalism discussed in the introduction with $\lambda_0$ playing the role of the rate function that determines when the next jump takes place, and \eqref{eq:Gill_where} implementing the transitions.}

\subsubsection{Simulation of the representation \eqref{eq:Gill_when}-\eqref{eq:Voltage2}}
\label{sec:sim_RTC_gill}

Simulation of \eqref{eq:Gill_when}-\eqref{eq:Voltage2} is the analog of using Gillespie's algorithm in the time-homogeneous case.  All random numbers generated in the algorithm below are assumed to be independent.

\vspace{.125in}

\begin{algorithm}[{For the simulation of the representation \eqref{eq:Gill_when}-\eqref{eq:Voltage2}}].
\label{gill-algorithm}
  \begin{enumerate}
  \item Initialize: %\footnote{Have to initialize the voltage also.  For any voltage there will be an equilibrium distribution of the channel states, and one could draw the initial channel state from this distribution if desired.}  
  set the initial number of molecules of each species, $X$. Set the initial voltage value $V$.  Set $t = 0$.  
  \item Let {$r$} be uniform(0,1)  and {numerically integrate} \eqref{eq:Voltage2} forward in time until 
  \[
  	\int_t^{t+\Delta}\lambda_0(V(s),X(s)){\, ds} = \ln(1/{r}).
\]
    \label{impstep3}
  \item Let $\xi$ be uniform(0,1) and find $k\in \{1,\dots,M\}$ for which 
  \[
  	\xi \in [q_{k-1}({(t+\Delta)-}),q_k({(t+\Delta)-})).
  \]
    \item Set $t = t + \Delta$ and 
    \[
    	X \leftarrow X +\zeta_k. 
    \]
    \label{modstep3}
  \item Return to step \ref{impstep3} or quit.
  \end{enumerate}
  \end{algorithm}

\bl{Note again that the above algorithm relies on us being able to calculate a hitting time.  If the relevant equations can be analytically solved for, then such numerical integration is unnecessary and efficiencies can be gained.}
%{The flowchart depicted in Figure \ref{fig:flowchart-gillespie} illustrates the algorithm \eqref{eq:Gill_when}-\eqref{eq:Voltage2} graphically.} {(DFA: again, I'm not a huge fan of the flowchart.)}
%\begin{figure}[htbp] %  figure placement: here, top, bottom, or page
%   \centering
%\includegraphics[height=8in]{GillespieAlgorithmFlowchart-v3.pdf}
%   \caption{{Flowchart illustrating one full iteration of the algorithm for simulation of \eqref{eq:Gill_when}-\eqref{eq:Voltage2}.  The algorithm is initialized by setting $t=0$ along with the values of $X(0)$ and $V(0)$.  The values $r$ and $\xi$ are drawn from independent random number streams. }}
%   \label{fig:flowchart-gillespie}
%\end{figure}

\section{Morris-Lecar}
\label{sec:ML}

%Our analysis applies to any conductance based model that admits a two dimensional slow manifold in its deterministic approximation (for example, Izhikevich's ``minimal'' persistent sodium + potassium spiking model \cite{Izhikevich2007}.) 

As a concrete illustration of the exact stochastic simulation algorithms, we will consider the well known Morris-Lecar system \cite{Rinzel+Ermentrout:1989}, developed as a model for oscillations observed in barnacle muscle fibers \cite{MorrisLecar1981BiophysJ}.
The deterministic equations, which correspond to an appropriate %the mean field or 
scaling limit of the system, constitute a planar model for the evolution of the membrane potential $v(t)$ and the fraction of potassium gates, $n\in[0,1]$, that are in the open or conducting state. In addition to a hyperpolarizing current carried by the potassium gates, there is a depolarizing calcium current gated by a rapidly equilibrating variable $m\in[0,1]$.  While a fully stochastic treatment of the Morris-Lecar system would include fluctuations in this calcium conductance, for simplicity {in this section} we will treat $m$ as a fast, deterministic variable in the same manner as in the standard {fast/slow decomposition, which we will refer to here as the planar Morris-Lecar model} \cite{ErmentroutTerman2010book}.  
{See Section \ref{sec:general} for a treatment of the Morris-Lecar system with both the potassium and calcium gates represented as discrete stochastic processes.}
%analysis \cite{ErmentroutTerman2010book}.

The deterministic or mean field equations {for the planar Morris-Lecar model} are:
\begin{eqnarray}\label{MLdet-v}
\frac{dv}{dt}&=&f(v,n)=\frac{1}{C}\left(I_{\text{app}}-g_{Ca}m_{\infty}(v)(v-v_{Ca}) -g_L(v-v_L)-g_Kn(v-v_K)\right)\\ \label{MLdet-n}
\frac{dn}{dt}&=&g(v,n)=\alpha(v)(1-n)-\beta(v)n
=(n_{\infty}(v)-n)/\tau(v)
\end{eqnarray}
{The kinetics of the potassium channel may be specified either by the instantaneous time constant and asymptotic target, $\tau$ and  $n_\infty$, or equivalently by the \textit{per capita} transition rates $\alpha$ and $\beta$.} 
The terms $m_{\infty},\alpha,\beta,n_\infty$ and $\tau$ satisfy 
\begin{eqnarray}
m_{\infty}&=&\frac{1}{2}\left(1+\tanh\left(\frac{v-v_a}{v_b}\right)\right)\\
\alpha(v)&=&\frac{\phi\cosh(\xi/2)}{1+e^{2\xi}}\\
\beta(v)&=&\frac{\phi\cosh(\xi/2)}{1+e^{-2\xi}}\\
n_{\infty}(v)&=&\alpha(v)/(\alpha(v)+\beta(v))=\left(1+\tanh\xi \right)/2\\
\tau(v)&=&1/(\alpha(v)+\beta(v))=1/\left(\phi\cosh(\xi/2)\right)
\end{eqnarray}
{where for convenience we define $\xi=(v-v_c)/v_d$.  }
For definiteness, we adopt values of the parameters 
\begin{eqnarray}
v_K&=&-84, v_L=-60,v_{Ca}=120\\
I_{\text{app}}&=&{100},g_K=8,g_L=2,C=20\\
v_a&=&-1.2,v_b=18\\
v_c&=&2,v_d=30,\phi=0.04,g_{Ca}=4.4
\end{eqnarray}
{for which the deterministic system has a stable limit cycle.  For smaller values of the applied current (e.g.~$I_\text{app}=75$) the system has a stable fixed point, that loses stability through a subcritical Hopf bifurcation as $I_\text{app}$ increases} (\cite{ErmentroutTerman2010book}, \S 3).

In order to incorporate the effects of random ion channel gating, we will introduce a finite number of potassium channels, $N_{\text{tot}}$, and treat the number of channels in the open state as a discrete random process, $0\le N(t)\le N_{\text{tot}}$.  In this simple model, each potassium channel switches between two states -- closed or open -- independently of the others, with voltage-dependent \textit{per capita}  transition rates $\alpha$ and $\beta$, respectively.  The entire population conductance ranges from $0$ to $g_K^oN_{\text{tot}}$, where $g_K^o=g_K/N_{\text{tot}}$ is the single channel conductance, and $g_K$ is the maximal whole cell conductance.  
%(Full expressions for $f$ and $g$ are found in Appendix \ref{app:ML}, however, we point out that $g$ is derived from the transition rates $\alpha$ and $\beta$, also given in the appendix.) 
%There are two gating variables representing different ion channels in the ML model.  To keep things as simple as possible, however, we will keep the usual deterministic interpretation of one of the variables (the $m$-gate, in this model representing a regenerative inward calcium current that leads to spike onset), and interpret only the slow variable ($n$, the potassium channel) as affected by channel noise from finite numbers of channels in a given cell.  Single channel conductances for potassium channels have been measured for (TBD: details) * and * and *, and range from *  to *.  Whole cell conductances range from * to *.  These measurements provide a wide range of possible numbers of potassium channels per cell in different cell and species types.  
For purposes of illustration and simulation we will typically use $N_{\text{tot}}=40$ individual channels.\footnote{Morris and Lecar used a value of $g_K=8\text{mmho/cm}^2$ for the specific potassium conductance, corresponding to 80 picoSiemens (pS) per square micron \cite{MorrisLecar1981BiophysJ}. The single channel conductance is determined by the structure of the potassium channel, and varies somewhat from species to species.  {However, conductances around 20 pS are typical \cite{ShingaiQuandt1986BrainRes}, which would give a density estimate of roughly 40 channels for a  10 square micron patch of cell membrane.}} 

\newcommand{\filt}{\mathcal{F}}
\newcommand{\bor}{\mathcal{B}}
\newcommand{\discrete}{\mathcal{P}}
\newcommand{\Yo}{Y_{\text{open}}}
\newcommand{\Yc}{Y_{\text{close}}}

Our random variables will therefore be the voltage, $V\in(-\infty,+\infty)$, and the number of open potassium channels, $N\in\{0,1,2,\cdots,N_{\text{tot}}\}$.  The number $N_{\text{tot}}\in\mathbb{N}$ is taken to be a fixed parameter.  We follow the usual capital/lowercase convention in the stochastic processes literature: $N(t)$ and $V(t)$ are the random processes and $n$ and $v$ are values they might take. {In the random time change representation of Section \ref{sec:RTC},} the opening and closing of the potassium channels are driven by two independent, unit rate Poisson processes, $\Yo(t)$ and $\Yc(t)$.%\footnote{Probability theoretic details: The underlying probability measures will be a product of Borel measure $\bor(\R)$ (for $V$) and the discrete measure $\discrete$ on $\mathbb{N}$ (for $N$).  With probability one, trajectories of $V$ will be piecewise differentiable, and trajectories for $N$ will be piecewise constant, right-continuous with left-hand limits (c\`{a}dl\`{a}g). The underlying filtration $\filt_t$ is that generated by the natural filtrations $\filt^o_t$ for $\Yo$, $\filt^c_t$ for $\Yc$, and the Brownian filtration $\filt^B_t$ for $V$, that is, $\filt_t=\filt^o_t\wedge\filt^c_t\wedge\filt^B_t$.  Is this correct?  Does it match the description in the preceding section?} 

The evolution of $V$ and $N$ is closely linked. Conditioned on $N$ having a specific value, say $N=n$, the evolution of $V$ obeys a \textit{deterministic} differential equation, 
\begin{equation}
\left.\frac{dV}{dt}\right|_{N=n}=f(V,n).
\end{equation}
Although its conditional evolution is deterministic, $V$ is nevertheless a random variable.
Meanwhile, $N$ evolves as a jump process, \textit{i.e.}~$N(t)$ is %almost-surely
 piecewise-constant, with transitions $N\to N\pm 1$ occurring with intensities that depend on the voltage $V$.  Conditioned on $V=v$, the transition rates for $N$ are
\begin{align}
N\to N+1 & \mbox{ with \textit{per capita} rate } \alpha(v)\\
\nonumber & \mbox{ (\textit{i.e.}~with net rate } \alpha(v)\cdot \left(N_{\text{tot}}-N\right)),\\
N\to N-1 & \mbox{ with \textit{per capita} rate } \beta(v)\\
\nonumber & \mbox{ (\textit{i.e.}~with net rate } \beta(v)\cdot N).
\end{align}
Graphically, we may visualize the state space for $N$ in the following manner:
\begin{equation*}
0 \overset{\alpha\cdot N_{\text{tot}}}{\underset{\beta}{\rightleftarrows}}1 
\overset{\alpha\cdot (N_{\text{tot}} - 1)}{\underset{2\beta}{\rightleftarrows}}2 
\cdots
(k-1)
\overset{\alpha\cdot (N_{\text{tot}} - k+1)}{\underset{k\beta}{\rightleftarrows}}k
\overset{\alpha\cdot (N_{\text{tot}} - k)}{\underset{(k+1)\beta}{\rightleftarrows}}(k+1)
\cdots
(N_{\text{tot}}-1)
\overset{\alpha}{\underset{N_{\text{tot}}\cdot \beta}{\rightleftarrows}}N_{\text{tot}},
\end{equation*}
with the nodes of the above graph being the possible states for the process $N$, and the transition intensities located above and below the transition arrows.
Adopting the random time change representation of Section \ref{sec:RTC} we write our stochastic Morris-Lecar system as follows (\textit{cf}.~equations (\ref{MLdet-v}-\ref{MLdet-n})):
\begin{align}\label{eq:ML-V}
	\frac{dV}{dt} &= f(V(t),N(t))=\\ \nonumber
	&=\frac{1}{C}\left(I_{\text{app}}-g_{Ca}m_{\infty}(V(t))(V(t)-V_{Ca}) {-g_L(V - V_L)- g_K^oN(t)(V(t)-V_K)}\right)\\	\label{eq:ML-N}
 \\ \nonumber
	N(t) &= N(0) - \Yc\left( \int_0^t \beta(V(s)) N(s) \, ds\right) + \Yo\left( \int_0^t \alpha(V(s)) (N_{\text{tot}} - N(s)) \, ds\right).
\end{align}

\vspace{.2in}

\begin{figure}[htbp] %  figure placement: here, top, bottom, or page
   \centering
   \includegraphics[width=6in]{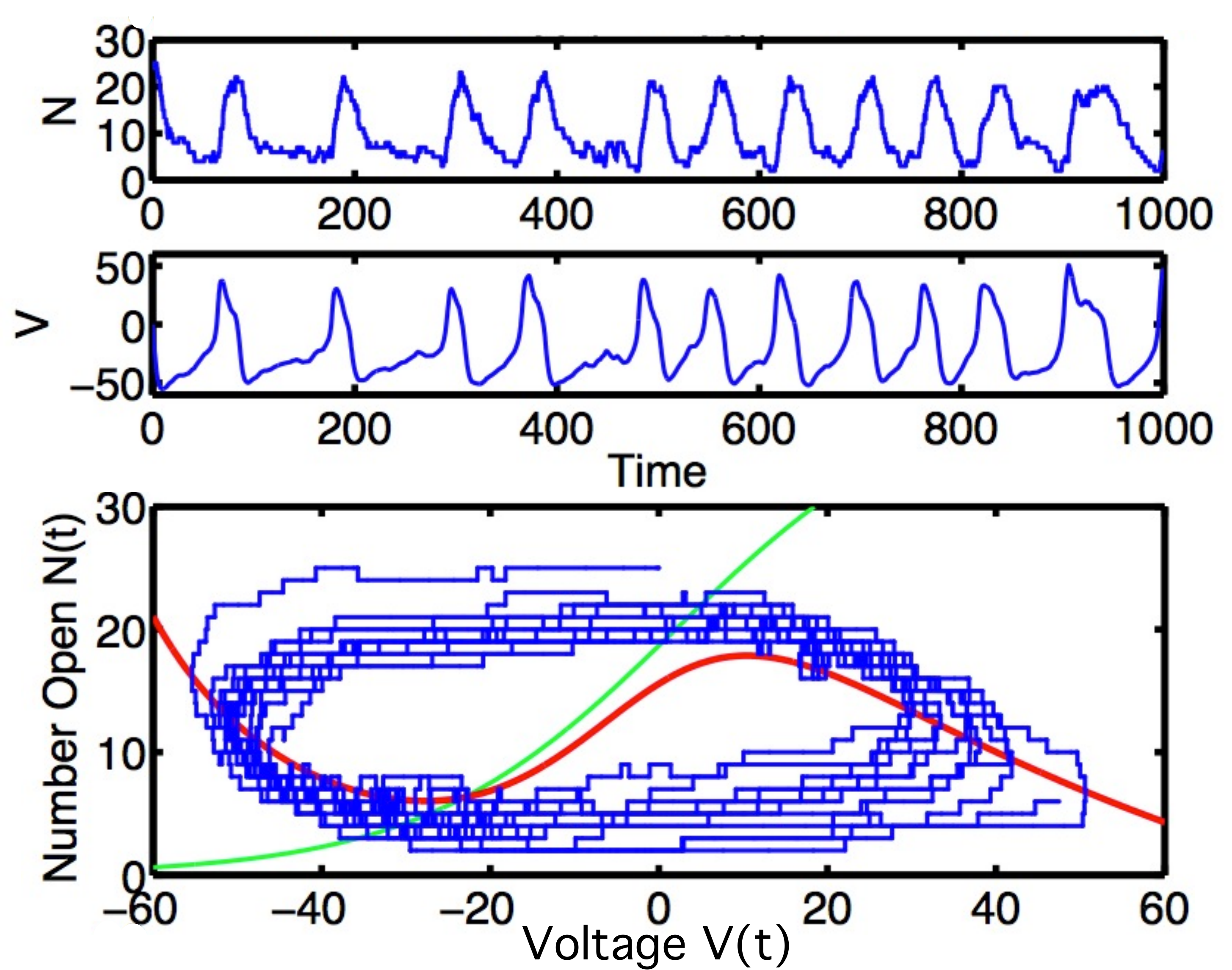} 
   \caption{Trajectory generated by Algorithm \ref{main-algorithm} (the random time change algorithm, \eqref{eq:RTC_main}-\eqref{eq:voltage}) for the planar Morris-Lecar model.  We set $N_{\text{tot}}=40$ potassium channels and used a driving current $I_{\text{app}}=100$, which is above the Hopf bifurcation threshold for the parameters given.  \textbf{Top Panel:} Number of open potassium channels ($N$), as a function of time.  \textbf{Second Panel:} Voltage ($V$), as a function of time.
   \textbf{Bottom Panel:} Trajectory plotted in the $(V,N)$ plane. 
   %{(DFA: I think this is the top panel.  I think the x-axis needs marking)}. 
   Voltage varies along a continuum while open channel number remains discrete. Red curve: $v$-nullcline of the underlying deterministic system, obtained by setting the RHS of equation (\ref{MLdet-v}) equal to zero.  Green curve: $n$-nullcline, obtained by setting the RHS of equation (\ref{MLdet-n}) equal to zero.}
   %{There seems to be a few things wrong with this caption.  Also, I cannot figure out what the red and green lines are.  Can you state what they are?}}
   \label{fig:ml-matlab-1channel}
\end{figure}

Appendices \ref{sec:xpp-k-only} and \ref{sec:matlab-k-only} provide sample implementations of {Algorithm \ref{main-algorithm}} for the planar Morris-Lecar equations in \texttt{XPP} and \texttt{Matlab}, respectively.  %{(DFA: Peter, can you clarify which algorithms are being implemented and/or can be found in the appendices?  Is it algorithm 1 (next reaction method), or Algorithm 2 (Gillespie)?)}
Figure \ref{fig:ml-matlab-1channel} illustrates the results of the \texttt{Matlab} implementation. 

%Next: include explicit statement of the exact Gillespie algorithm.  And samples from simulations.  And a pointer to the full code in an appendix or two. 

\section{Models with more than one channel type}
\label{sec:general}

\bl{We present here an example of a conductance based model with more than one stochastic gating variable.  Specifically, we consider the original Morris-Lecar model, which has a three dimensional phase space, and we take both the calcium and potassium channels  to be discrete.  We include code in Appendices \ref{sec:xpp-both} and \ref{sec:matlab-both}, for \texttt{XPP} and \texttt{Matlab}, respectively,  for the implementation of Algorithm \ref{main-algorithm} for this model.}

%Of course, the algorithms presented above readily generalize to systems with more than one type of stochastic channel, such as the Hodgkin-Huxley system \cite{jp:Hodgkin+Huxley:1952d,SkaugenWalloe1979ActaPhysiolScand}.  As an illustration, we include code in Appendices \ref{sec:xpp-both} and \ref{sec:matlab-both}, for \texttt{XPP} and \texttt{Matlab}, respectively, implementing a stochastic version of the original Morris-Lecar model, which has a three dimensional phase space with both the calcium and potassium channels taken to be discrete.  , using the random time change algorithm (\eqref{eq:RTC_main}-\eqref{eq:voltage}). % {(DFA: Peter, can you clarify if the implementations are Algorithm 1 or 2?)}

\subsection{Random Time Change Representation for the Morris-Lecar Model with Two Stochastic Channel Types}
\label{sec:ML2}

In Morris and Lecar's original treatment of voltage oscillations in barnacle  muscle fiber \cite{MorrisLecar1981BiophysJ} the calcium gating variable $m$ is included as a dynamical variable.  The full (deterministic) equations have the form:
\begin{eqnarray}\label{MLdet-v-full}
\frac{dv}{dt}&=&F(v,n,m)=\frac{1}{C}\left(I_{\text{app}} -g_L(v-v_L)-g_{Ca}m(v-v_{Ca})-g_Kn(v-v_K)\right)\hspace{10mm}\\ \label{MLdet-n-full}
\frac{dn}{dt}&=&G(v,n,m)=\alpha_n(v)(1-n)-\beta_n(v)n
=(n_{\infty}(v)-n)/\tau_n(v)\\ \label{MLdet-m-full}
\frac{dm}{dt}&=&H(v,n,m)=\alpha_m(v)(1-m)-\beta_m(v)m
=(m_{\infty}(v)-m)/\tau_m(v)
\end{eqnarray}
{Here, rather than setting $m$ to its asymptotic value $m_\infty=\alpha_m/(\alpha_m+\beta_m)$, we allow the number of calcium gates to evolve according to \eqref{MLdet-m-full}.}  
The planar form (Equations \ref{MLdet-v}-\ref{MLdet-n}) is obtained by observing that $m$ approaches equilibrium significantly more quickly than $n$ and $v$.  Using standard arguments from singular perturbation theory \cite{Rinzel+Ermentrout:1989,Rubin+Terman:2002}, one may approximate certain aspects of the full system \eqref{MLdet-v-full}-\eqref{MLdet-m-full} by setting $m$ to $m_\infty(v)$, and replacing $F(v,n,m)$ and $G(v,n,m)$ with $f(v,n)=F(v,n,m_\infty(v))$ and $g(v,n)=G(v,n,m_\infty(v))$, respectively.  {This reduction to the slow dynamics leads to the planar model \eqref{MLdet-v}-\eqref{MLdet-n}.}

{To specify the full 3D equations, we introduce $\xi_m=(v-v_a)/v_b$ in addition to $\xi_n=(v-v_c)/v_d$ already introduced for the 2D model.  
The variable $\xi_x$ represents where the voltage falls along the activation curve for channel type $x$, relative to its half-activation point ($v_a$ for calcium and $v_c$ for potassium) and its slope (reciprocals of $v_b$ for calcium and $v_d$ for potassium).}
The per capita opening rates $\alpha_m$, $\alpha_n$ and closing rates $\beta_m$, $\beta_n$ for each channel type are given by
\begin{eqnarray}
\alpha_m(v)&=&\frac{\phi_m\cosh(\xi_m/2)}{1+e^{2\xi_m}},\,\,\,
\beta_m(v)=\frac{\phi_m\cosh(\xi_m/2)}{1+e^{-2\xi_m}}\\
\alpha_n(v)&=&\frac{\phi_n\cosh(\xi_n/2)}{1+e^{2\xi_n}},\,\,\,
\beta_n(v)=\frac{\phi_n\cosh(\xi_n/2)}{1+e^{-2\xi_n}}
\end{eqnarray}
with parameters $v_a=-1.2, v_b=18, v_c=2, v_d=30, \phi_m=0.4, \phi_n=0.04$.
{The asymptotic open probabilities for calcium and potassium are given, respectively, by the terms $m_{\infty},n_\infty$, and the time constants by $\tau_m$ and $\tau_n$.  These terms satisfy the relations} 
\begin{eqnarray}
m_{\infty}(v)&=&\alpha_m(v)/(\alpha_m(v)+\beta_m(v))=\left(1+\tanh \xi_m\right)/2\\
n_{\infty}(v)&=&\alpha_n(v)/(\alpha_n(v)+\beta_n(v))=\left(1+\tanh\xi_n \right)/2\\
\tau_m(v)&=&1/\left(\phi\cosh(\xi_m/2)\right)\\
\tau_n(v)&=&1/\left(\phi\cosh(\xi_n/2)\right).
\end{eqnarray}
{Assuming a finite population of $M_{\text{tot}}$ calcium gates and $N_{\text{tot}}$ potassium gates, \bl{we have a stochastic hybrid system with one continuous variable, $V(t)$, and two discrete variables, $M(t)$ and $N(t)$.}  The voltage evolves as according to the sum of the applied, leak, calcium, and potassium currents:} 
\begin{eqnarray}
\frac{dV}{dt}&=&F(V(t),N(t),M(t))\\
&=&\frac{1}{C}\left(I_{\text{app}} -g_L(V(t)-v_L)-g_{Ca}\frac{M(t)}{M_\text{tot}}(V(t)-v_{Ca})-g_K\frac{N(t)}{N_{\text{tot}}}(V(t)-v_K)\right), \label{MLstoch-both-full} \nonumber
\end{eqnarray}
while the number of open $M$ and $N$ remain constant except for unit increments and decrements.
\bl{The discrete channel states $M(t)$ and $N(t)$ evolve according to
\begin{align}
	M(t) &= M(0) - \Yc\left( \int_0^t \beta_m(V(s)) M(s) \, ds\right) + \Yo\left( \int_0^t \alpha_m(V(s)) (M_{\text{tot}} - M(s)) \, ds\right)\\
	N(t) &= N(0) - \Yc\left( \int_0^t \beta_n(V(s)) N(s) \, ds\right) + \Yo\left( \int_0^t \alpha_n(V(s)) (N_{\text{tot}} - N(s)) \, ds\right).
\end{align}
}

\begin{figure}[t]%[htbp] %  figure placement: here, top, bottom, or page
   \begin{center}
   \includegraphics[width=3.2in]{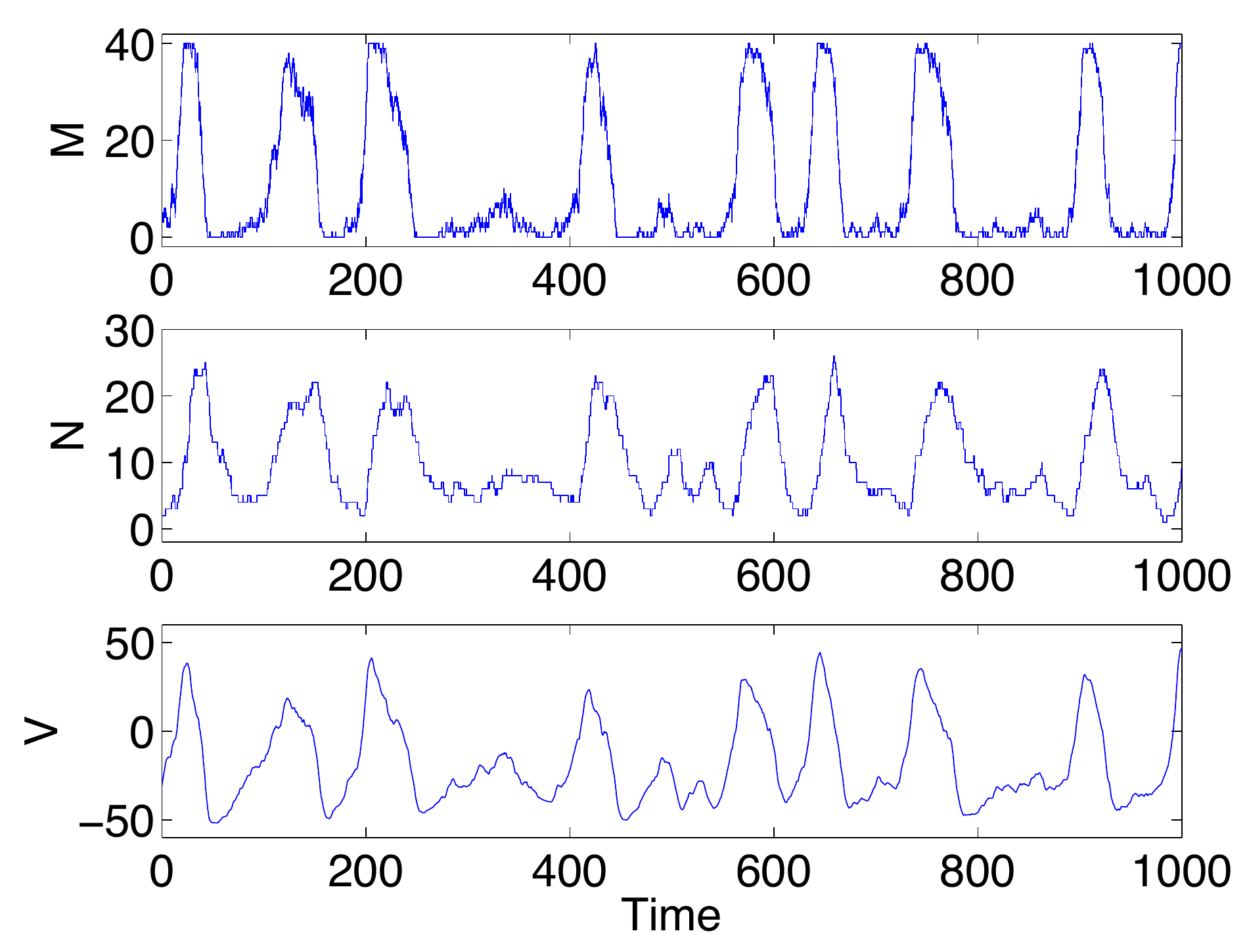} 
   \includegraphics[width=3.2in]{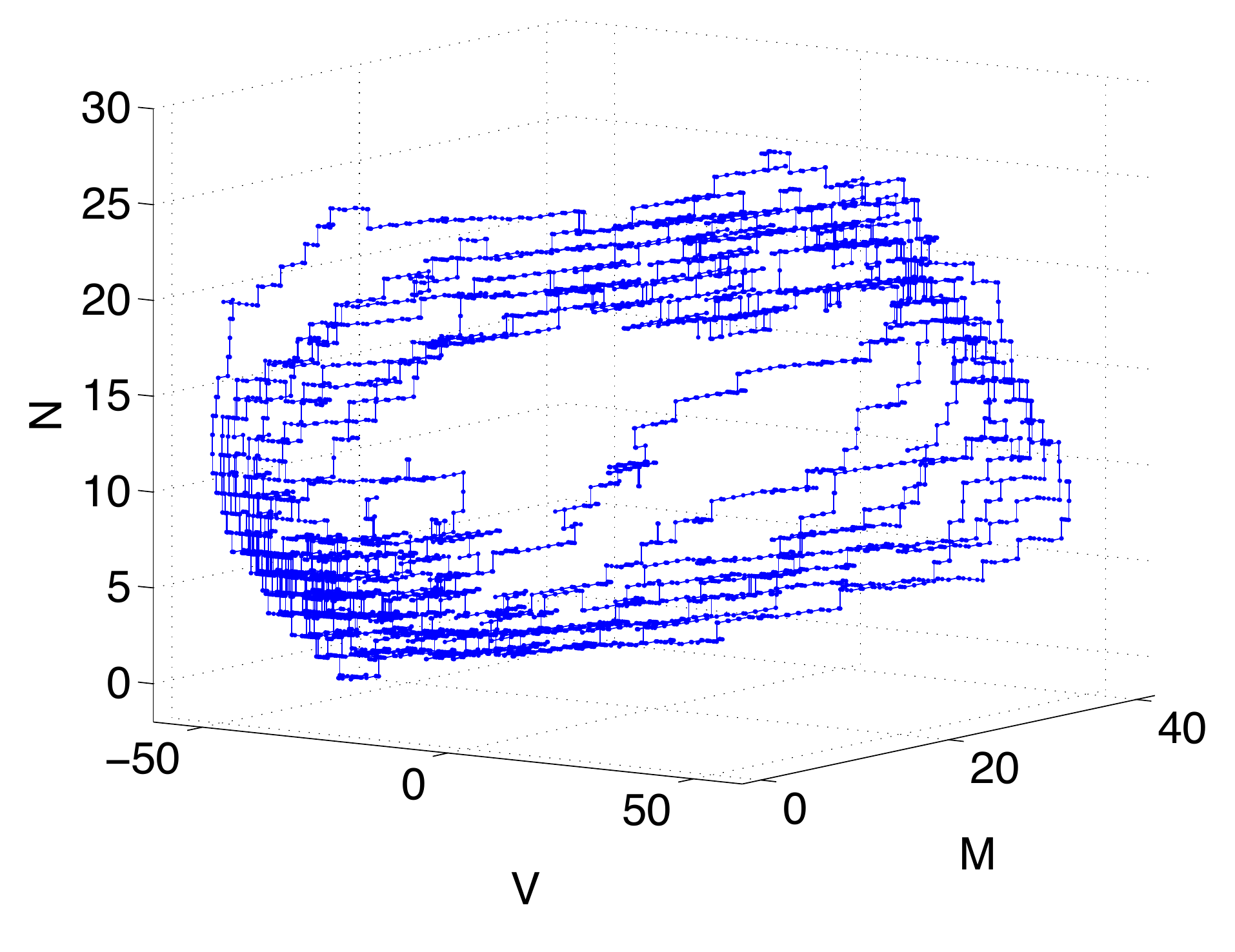}
   \end{center}
   \caption{Trajectory generated by Algorithm \ref{main-algorithm} (the random time change algorithm, 
   %Illustration {generated by Algorithm \ref{main-algorithm}} for simulation of 
   \eqref{eq:RTC_main}-\eqref{eq:voltage}) for the full three-dimensional Morris-Lecar model (equations \eqref{MLdet-v-full}-\eqref{MLdet-m-full}). 
   %{(DFA: shouldn't there be more equations?)}).  
   We set $N_{\text{tot}}=40$ potassium channels and $M_{\text{tot}}=40$ calcium channels, and used a driving current $I_{\text{app}}=100$, {a value} above the Hopf bifurcation threshold of the mean field equations for the parameters given. \textbf{Top {Left} Panel:} Number of open calcium channels ($M$), as a function of time. \textbf{Second {Left} Panel:} Number of open potassium channels ($N$), as a function of time.  \textbf{Third {Left} Panel:} Voltage ($V$), as a function of time.
   \textbf{{Right} Panel:} Trajectory plotted in the $(V,M,N)$ phase space. Voltage varies along a continuum while the joint channel state remains discrete.  
   Note that the number of open calcium channels makes frequent excursions between $M=0$ and $M=40$, which demonstrates that neither a Langevin approximation nor an approximate algorithm such as $\tau$-leaping (Euler's method) would provide a good approximation to the dynamics of the system. 
   % invalidating the {assumption of approximately Gaussian increments in channel state underlying the common chemical Langevin or ``$\tau$-leaping" approximation \cite{CaoGillespiePetzold2006JChemPhys,Gillespie2007AnnRevPhysChem} in this parameter range.} {(DFA: I'm not sure what the Gaussian tau-leaping algorithm is.  Do we mean using Euler-Maruyama on the diffusion equations?)}
   } 
   \label{fig:ml-matlab-both-channels}
\end{figure}
{Figure \ref{fig:ml-matlab-both-channels} shows the results of the Matlab implementation for the 3D Morris-Lecar system, with both the potassium and calcium channel treated discretely, using Algorithm \ref{main-algorithm} (the random time change algorithm, \eqref{eq:RTC_main}-\eqref{eq:voltage}).  Here $M_{\text{tot}}=N_{\text{tot}}=40$ channels, and the applied current $I_{\text{app}}=100$ puts the deterministic system at a stable limit cycle close to a Hopf bifurcation.}

\section{Comparison of the Exact Algorithm with a Piecewise Constant Propensity Approximation.}
\label{sec:convergence}
Exact versions of the stochastic simulation algorithm for hybrid ion channel models have been known since at least the 1980s \cite{ClayDeFelice1983BiophysJ}.  Nevertheless, the implementation one finds most often used in the literature is an approximate method in which the per capita reaction propensities are held fixed between channel events.  
%(e.g.~\cite{FischSchwalgerLindnerHerzBenda2012JNSci,KisperskyTilman2008Scholarpedia,SchwalgerFischBendaLindner2010PLosCB}).  
That is, in 
step 3 of {Algorithm \ref{main-algorithm}} the integral
$$\int_t^{t+\Delta_k}\lambda_k(V(s),X(s))\,ds$$
is replaced with 
$$\Delta_k\,\lambda_k(V(t),X(t))$$
leaving the remainder of the algorithm  unchanged.  Put another way, one generates the sequence of channel state jumps using the propensity immediately following the most recent jump, rather than taking into account the time dependence of the reaction propensities due to the continuously changing voltage.  This piecewise constant propensity approximation is  analogous, in a sense, to the forward Euler  method for the numerical solution of ordinary differential equations.  %{\sout{It is a ``forward" approximation in that the rate of change of the hazard function of the jump Markov process is held fixed at the value determined at the beginning of the previous time step.  The method is distinct, of course, in that the time steps are random, depending on the placement of the points of the underlying Poisson processes, rather than being fixed by a preset step size.}} }} ({DAVE:  I just don't think the last bit is needed. It seemed clear to me.})%{(DFA: sort of.  Forward Euler typically has a fixed step-size, whereas this approximate algorithm has a ``step size'' which is random and depends upon the placement of the points of the Poisson process.)}  } 

\begin{figure}[h!] %  figure placement: here, top, bottom, or page
   \centering
   \includegraphics[width=6in]{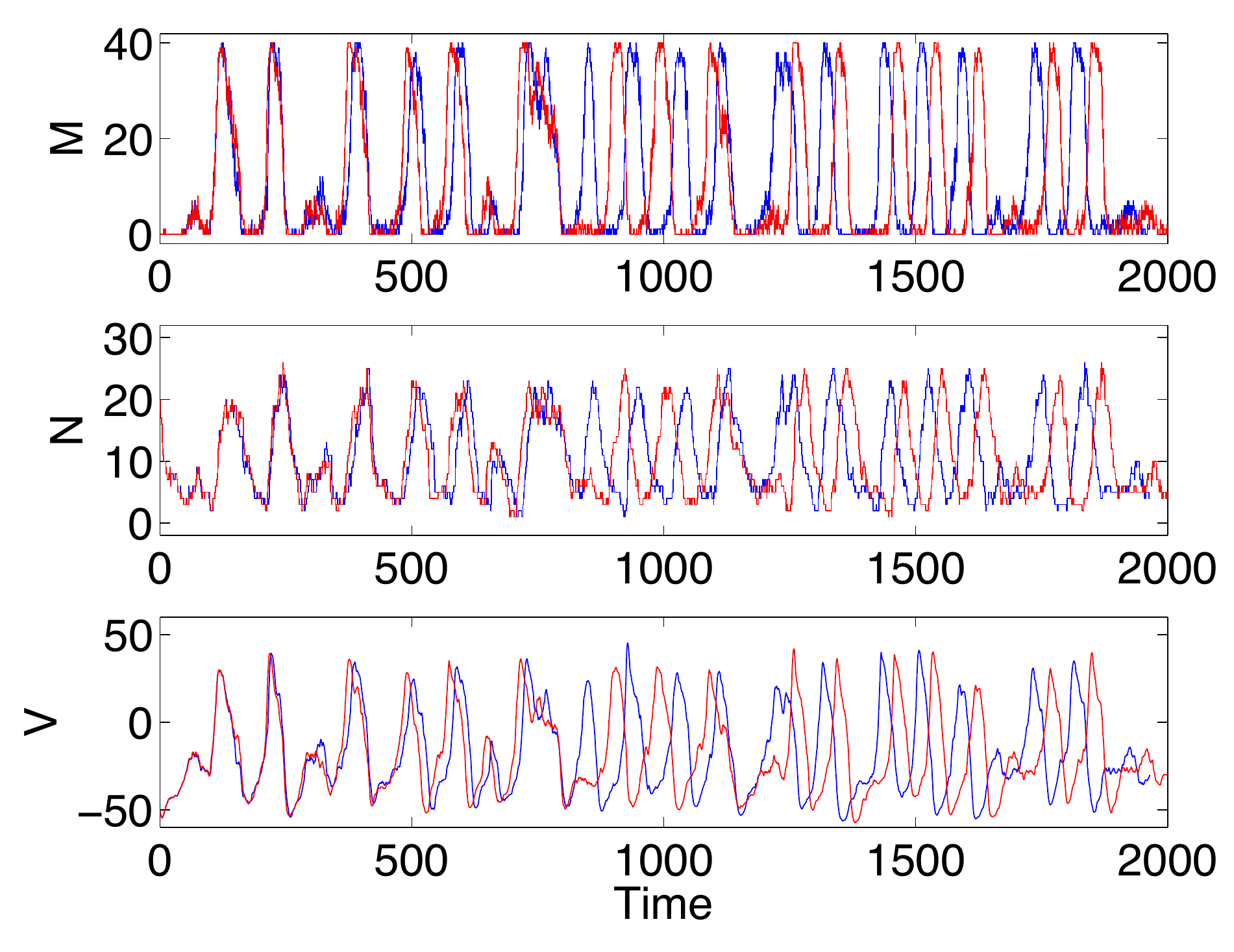} 
   \caption{{Comparison of the exact algorithm with the %``forward" 
   piecewise constant propensity  approximation.   Blue solid lines denote the solution $(M(t),N(t),V(t))$ obtained using the Algorithm \ref{main-algorithm}.  Red dashed lines denote the solution $(\tilde{M}(t),\tilde{N}(t),\tilde{V}(t))$ obtained using the \bl{piecewise constant approximation}. Both algorithms were begun with identical initial conditions $(M_0,N_0,V_0)$, and driven by the same four Poisson process streams $Y_1,\cdots,Y_4$.
   Note the gradual divergence of the trajectories as differences between the exact and the forward approximate algorithms accumulate, demonstrating ``strong" (or pathwise) divergence of the two methods.  
   The exact and approximate trajectories diverge as time increases, even though they are driven by identical noise sources.  %using data from 
   %projects/aplysia/code/channelnoise/matlab/mlexactstrongcompare_Iapp100_M40_N40
   %projects/aplysia/code/channelnoise/matlab/mlexactcompare.m
   }  
   }
   \label{fig:compare}
\end{figure}

Figure \ref{fig:compare} shows a direct comparison of pathwise numerical solutions obtained by the exact method provided in Algorithm \ref{main-algorithm} and the approximate \bl{piecewise constant method detailed above}. In general, the solution of a stochastic differential equation {with a given initial condition} is a map from the sample space $\Omega$ %{\sout{, as well as the initial conditions,}} 
to a space of trajectories.  In the present context, the underlying sample space consists of one independent unit rate Poisson process per reaction channel.   For the planar Morris-Lecar model a point in $\Omega$ amounts to fixing two Poisson processes, $Y_{\text{open}}$ and $Y_{\text{closed}}$, to drive the transitions of the potassium channel.  For the full 3D Morris-Lecar model we have four processes, $Y_1\equiv Y_{\text{Ca,open}}, Y_2\equiv Y_{\text{Ca,closed}}, Y_3\equiv Y_{\text{K,open}}$ and $Y_4\equiv Y_{\text{K,closed}}$.  In this case the exact algorithm provides a numerical  solution of the map from $\{Y_k\}_{k=1}^4\in\Omega$ and initial conditions $(M_0,N_0,V_0)$ to the trajectory $(M(t),N(t),V(t))$.  The approximate piecewise constant algorithm gives a map from the same domain to a  different trajectory, $(\tilde{M}(t),\tilde{N}(t),\tilde{V}(t))$.  To make a pathwise comparison for the full Morris-Lecar model, therefore, we fix both the initial conditions and the four Poisson processes, and compare the resulting trajectories.  

Several features are evident in Figure \ref{fig:compare}.  Both algorithms produce a sequence of noise-dependent voltage spikes, with similar firing rates.  The  trajectories $(M,N,V)$ and $(\tilde{M},\tilde{N},\tilde{V})$ initially remain close together, and the timing of the first spike (taken, e.g., as an upcrossing of $V$ from negative to positive voltage) is similar for both algorithms.  Over time, however,  discrepancies between the trajectories accumulate.  The timing of the second and third spikes is noticeably different, and before ten spikes have accumulated the spike trains have become effectively uncorrelated.

Even though trajectories generated by the exact and approximate algorithms diverge when driven by identical Poisson processes, the two processes could still generate sample paths with similar time-dependent or stationary distributions.  That is, even though the two algorithms show strong divergence, they could still {be close in a weak sense.

Given $M_{tot}$ calcium and $N_{tot}$ potassium channels, the density for the hybrid Markov process %{\sout{(and the sample paths of either version of the exact algorithm)}}
may be written 
\begin{equation}
\rho_{m,n}(v,t)=\frac{1}{dv}\Pr\left\{M(t)=m,N(t)=n,V\in[v,v+dv)\right\},
\end{equation}
and obeys a master equation 
\begin{align}\label{eq:Kolmogorov}
\frac{\partial \rho_{m,n}(v,t)}{\partial t} =&-\frac{\partial \left( F(v,n,m)  \rho_{m,n}(v,t) \right)  }{\partial v}\\
\nonumber
&- \left(\alpha_m(v)(M_{\text{tot}}-m)+\beta_m(v)m+\alpha_n(v)(N_{\text{tot}}-n)+\beta_n(v)n \right)\rho_{m,n}(v,t)
\\ 
\nonumber
&+(M_{\text{tot}}-m+1)\alpha_m(v)\rho_{m-1,n}(v,t)+(m+1)\beta_m(v)\rho_{m+1,n}(v,t)\\
\nonumber
&+(N_{\text{tot}}-n+1)\alpha_n(v)\rho_{m,n-1}(v,t)+(n+1)\beta_n(v)\rho_{m,n+1}(v,t),
\end{align}
with initial condition $\rho_{m,n}(v,0)\ge 0$ given by any (integrable) density such that
$\int_{v\in\R}\sum_{m,n}\rho_{m,n}(v,0)\,dv\equiv 1$, and boundary conditions $\rho\to 0$ as $|v|\to\infty$ and $\rho\equiv 0$ for either $m,n<0$ or $m>M_{\text{tot}}$ or $n>N_{\text{tot}}$.

In contrast, the approximate algorithm with piecewise constant propensities does not generate a Markov process, since the transition probabilities depend on the past rather than the present values of the voltage component.  Consequently they do not satisfy a master equation. Nevertheless it is plausible that they may have a unique stationary distribution.  

{Figure \ref{fig:v+n-histograms} shows pseudocolor plots of the histograms viewed in the $(v,n)$ plane, i.e.~with entries summed over $m$, for Algorithm \ref{main-algorithm} (``Exact") %exact {(DFA: which exact algorithm is being used, 1 or 2?)} 
and the \bl{approximate piecewise constant (``PC") algorithm}, with $M_{\text{tot}}=N_\text{tot}=k$ channels for $k=1,2,5,10,20$ and $40$.  The two algorithms were run with independent random number streams in the limit cycle regime (with $I_\text{app}=100$) for $t_\text{max}\approx 200,000$ time units, sampled every 10 time units, which generated $\ge 17,000$ data points per histogram.  For $k<5$ the difference in the histograms is obvious at a glance.  For $k\ge 10$ the histograms appear increasingly similar.  }  %{Dave: I'm still confused.  Do we actually use $d(\rho,\tilde \rho)$?  It seems that this paragraph implies you are plotting the amount of time the process spends in different states.  This is not $d(\rho,\tilde \rho)$.}

{Figure \ref{fig:v-histograms} shows bar plots of the histograms \bl{with 2,000,000 sample points} projected on the voltage axis, i.e.~with entries summed over $m$ and $n$, for the same data as in Figure \ref{fig:v+n-histograms}, with $M_{\text{tot}}=N_\text{tot}=k$ channels ranging from $k=1$ to \bl{$k=100$}.  Again, for $k\le 5$, the two algorithms generate histograms that are clearly distinct.  For $k\ge 20$ they appear similar, while $k=10$ appears to be a borderline case.  }

\begin{figure}[htbp] %  figure placement: here, top, bottom, or page
   \centering
   \includegraphics[width=3.2in]{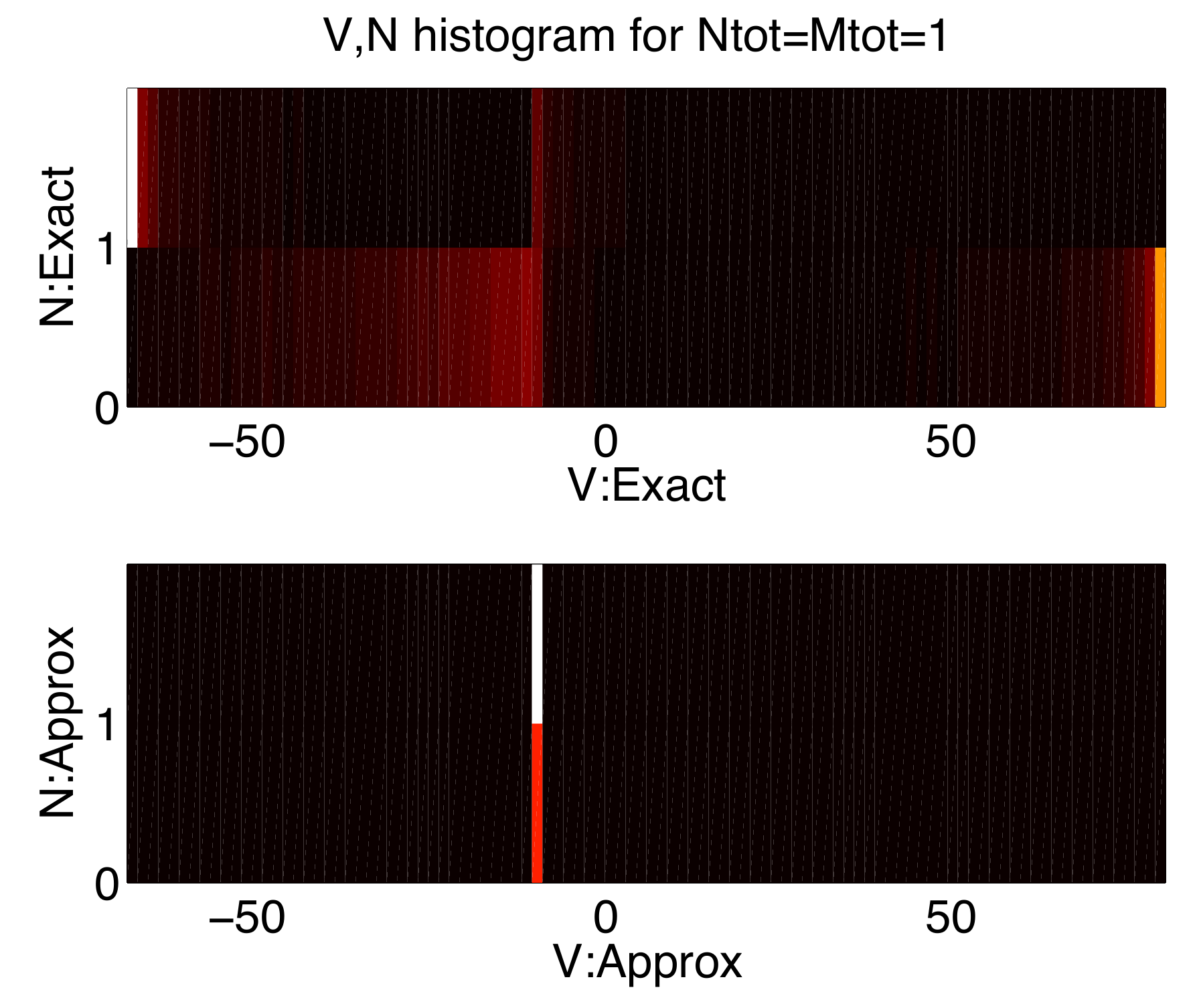} 
   \includegraphics[width=3.2in]{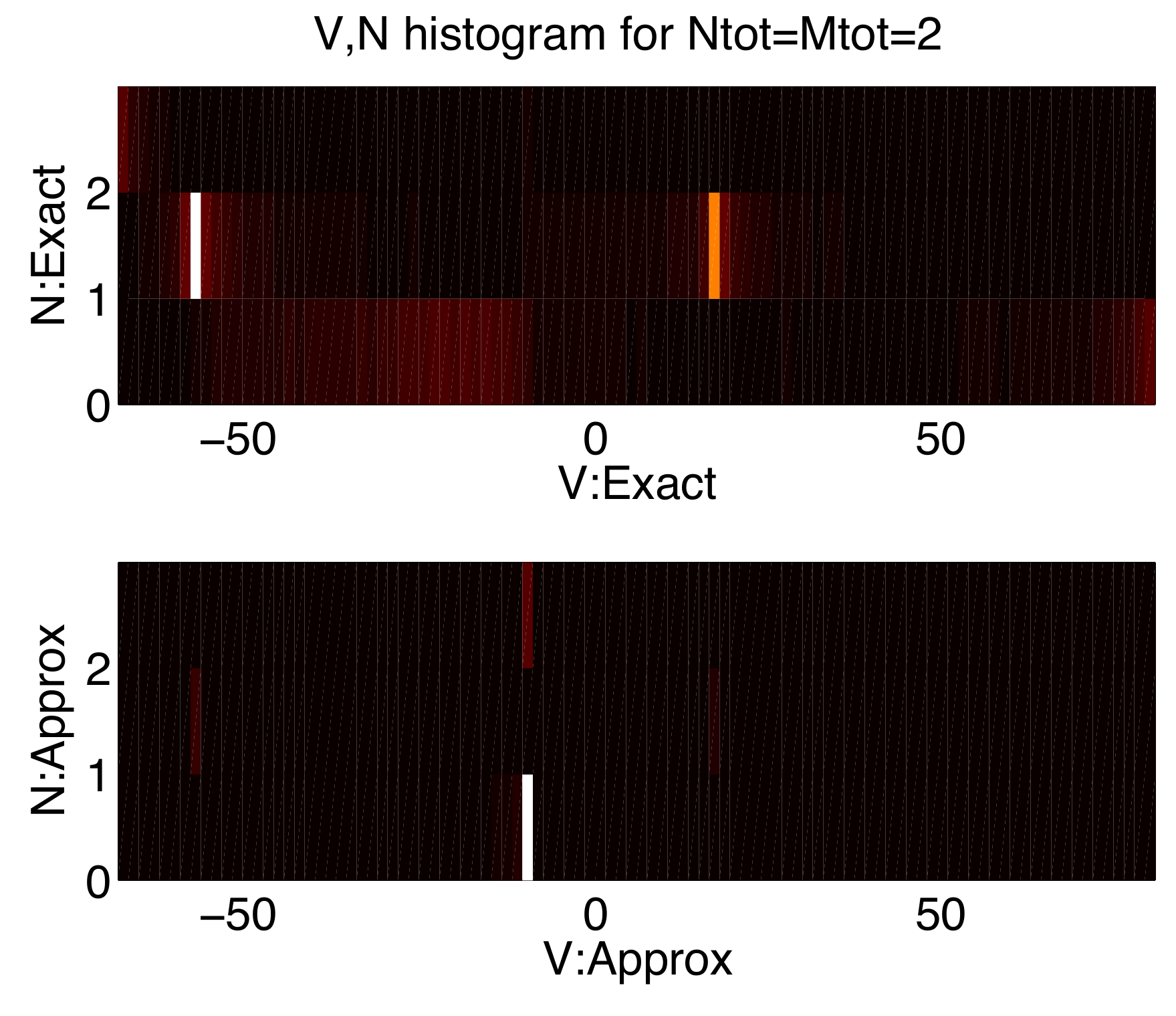} 
   \includegraphics[width=3.2in]{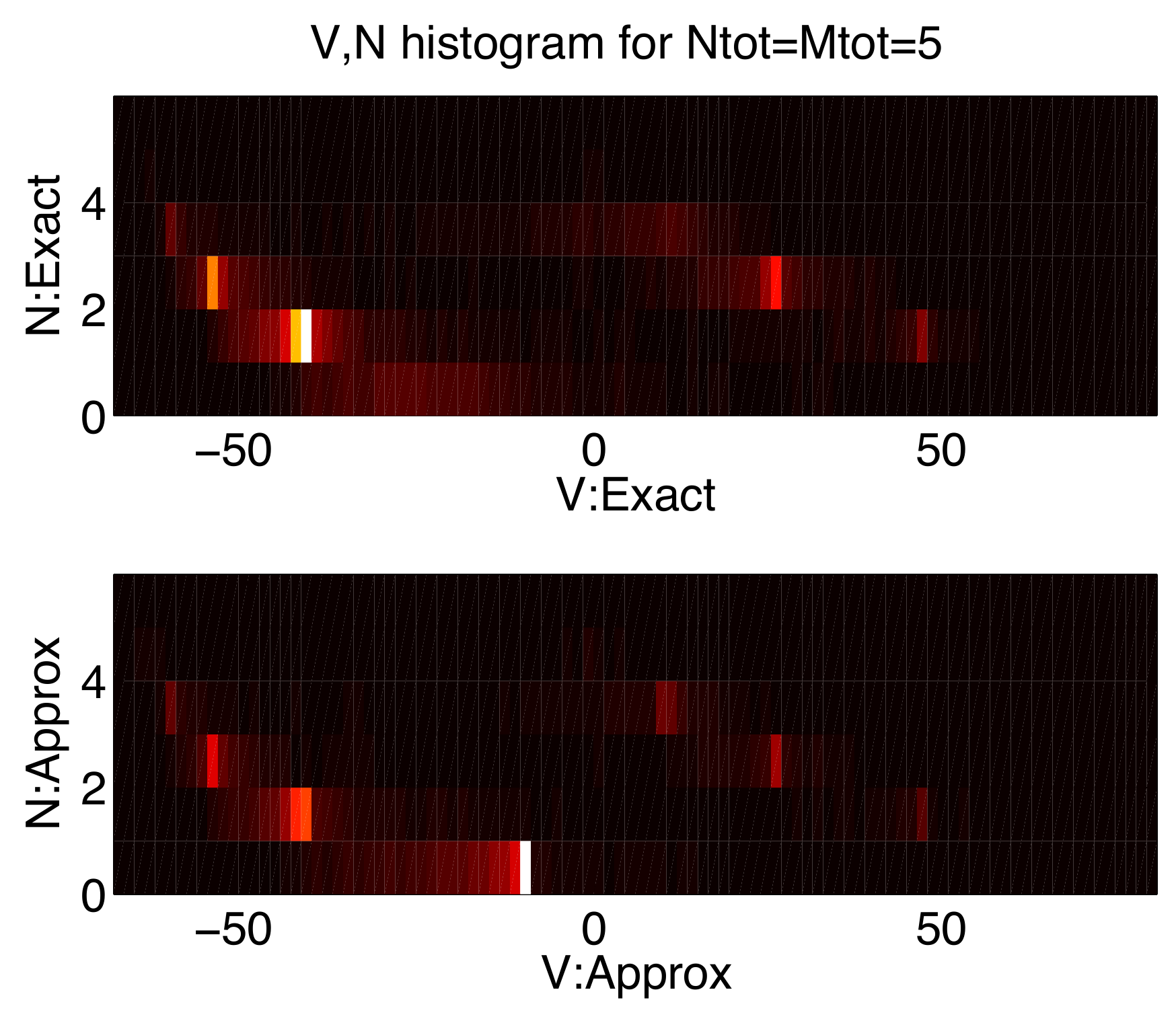} 
   \includegraphics[width=3.2in]{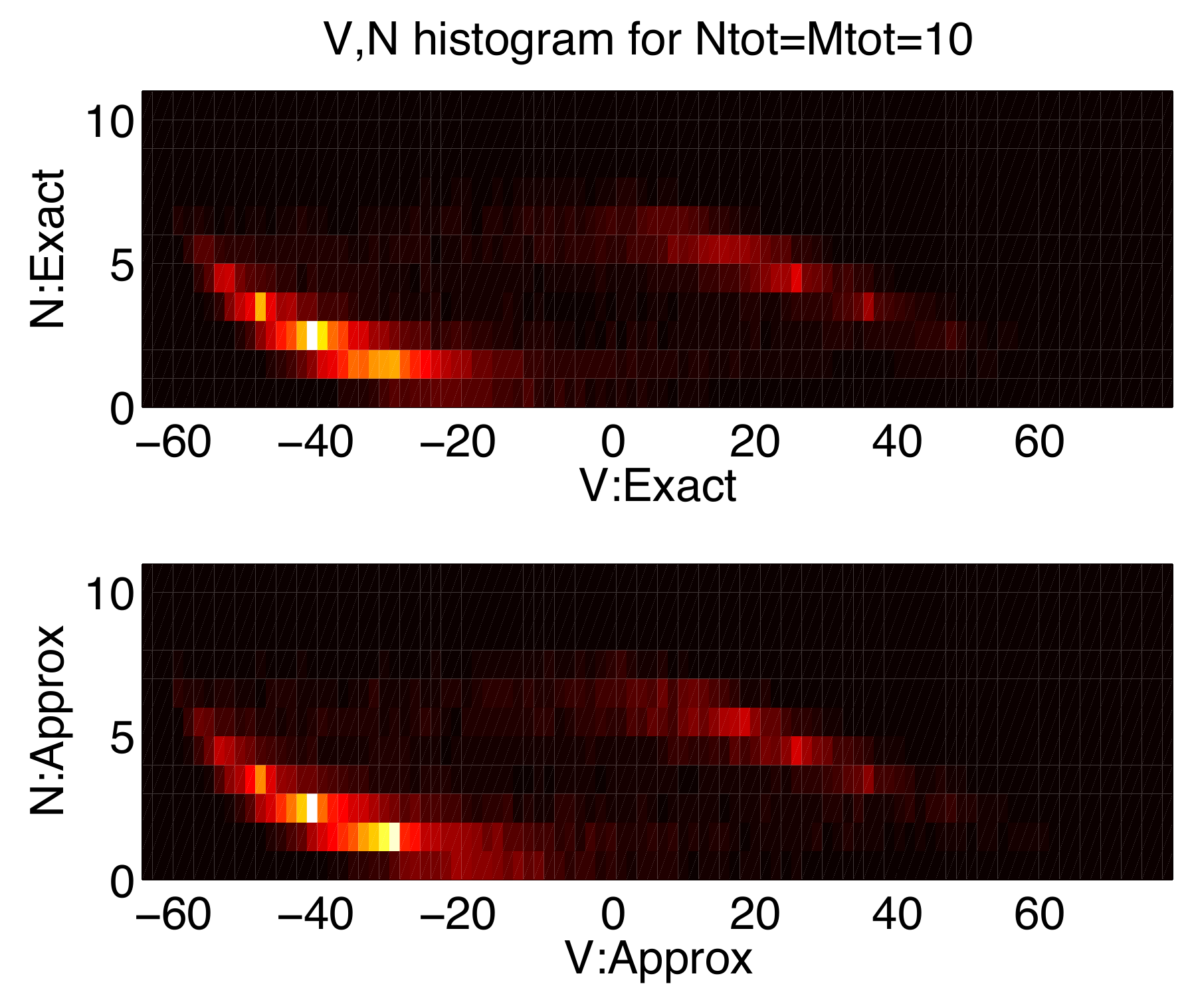} 
   \includegraphics[width=3.2in]{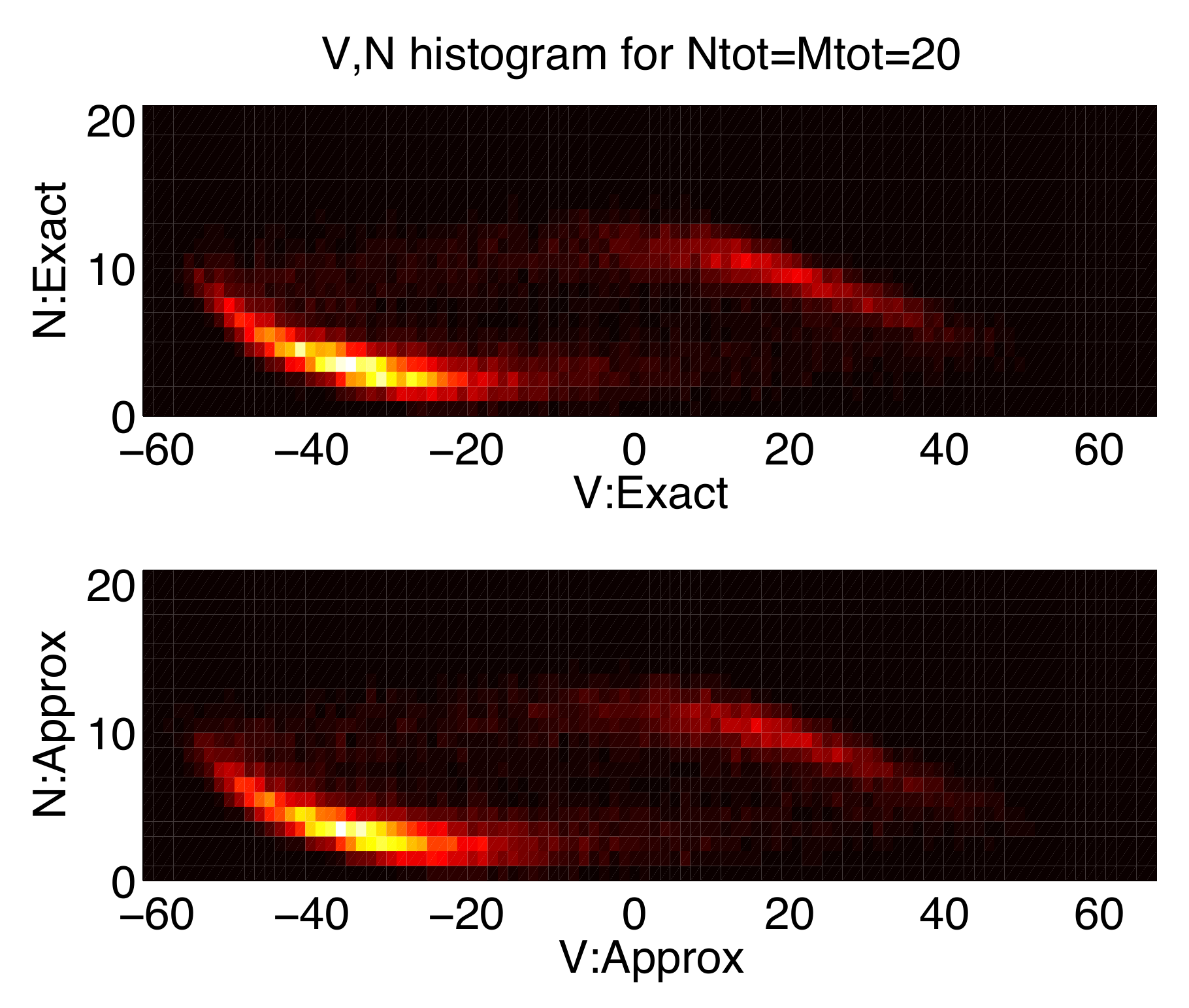} 
   \includegraphics[width=3.2in]{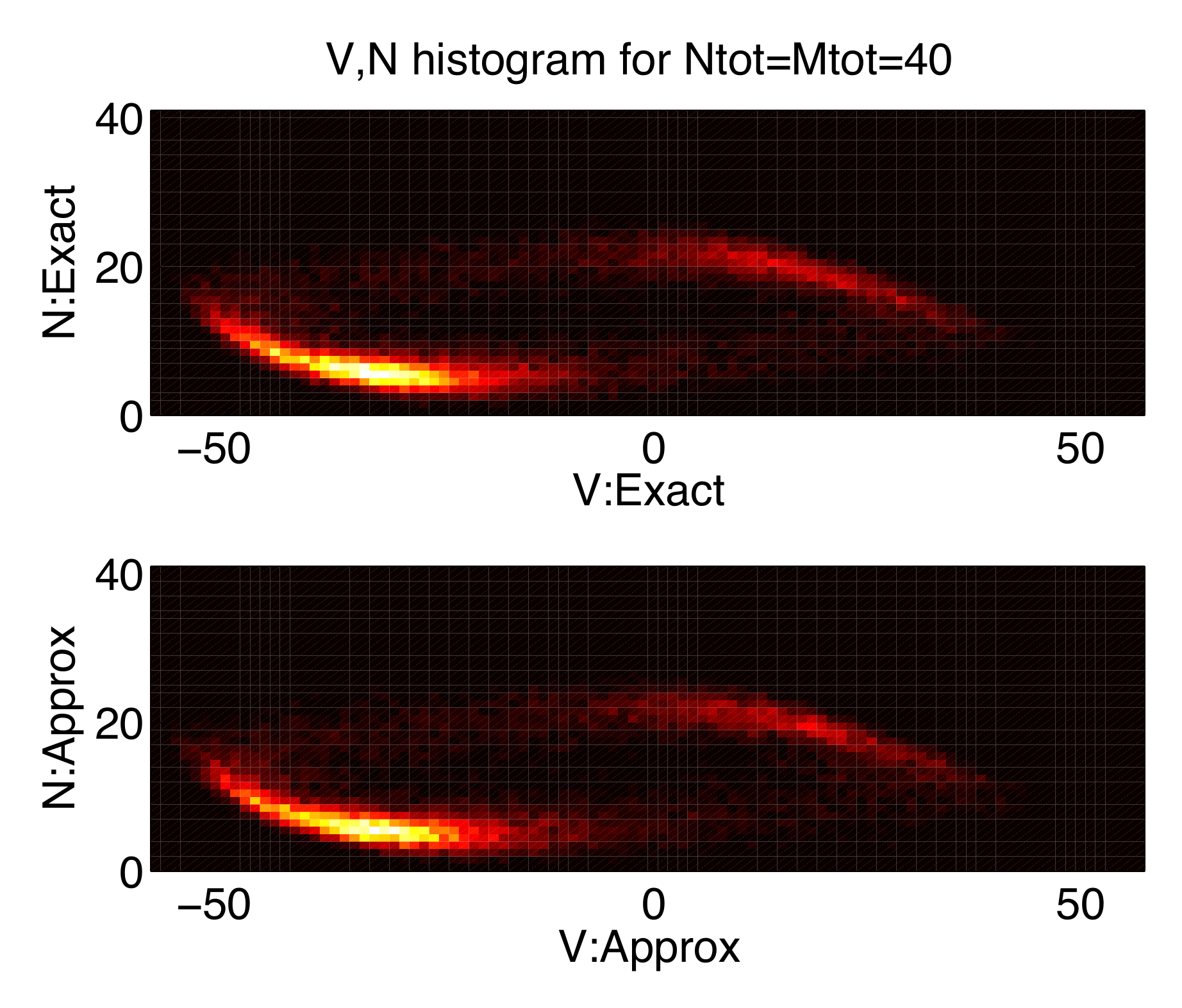} 
   \caption{Normalized histograms of voltage and $N$ for exact and \bl{approximate piecewise constant} algorithms. The $V$-axis was partitioned into 100 equal width bins for each pair of histograms.  The $N$-axis takes on $N_{\text{tot}}+1$ discrete values.  Color scale indicates relative frequency with which a bin was occupied (dark = infrequent, lighter=more frequent).  %{(DFA: Can you say what the colors mean?  I'm having trouble figuring out what this plot tells me)}
   }
   \label{fig:v+n-histograms}
\end{figure}

\begin{figure}[htbp] %  figure placement: here, top, bottom, or page
   \centering
   \includegraphics[width=3.2in]{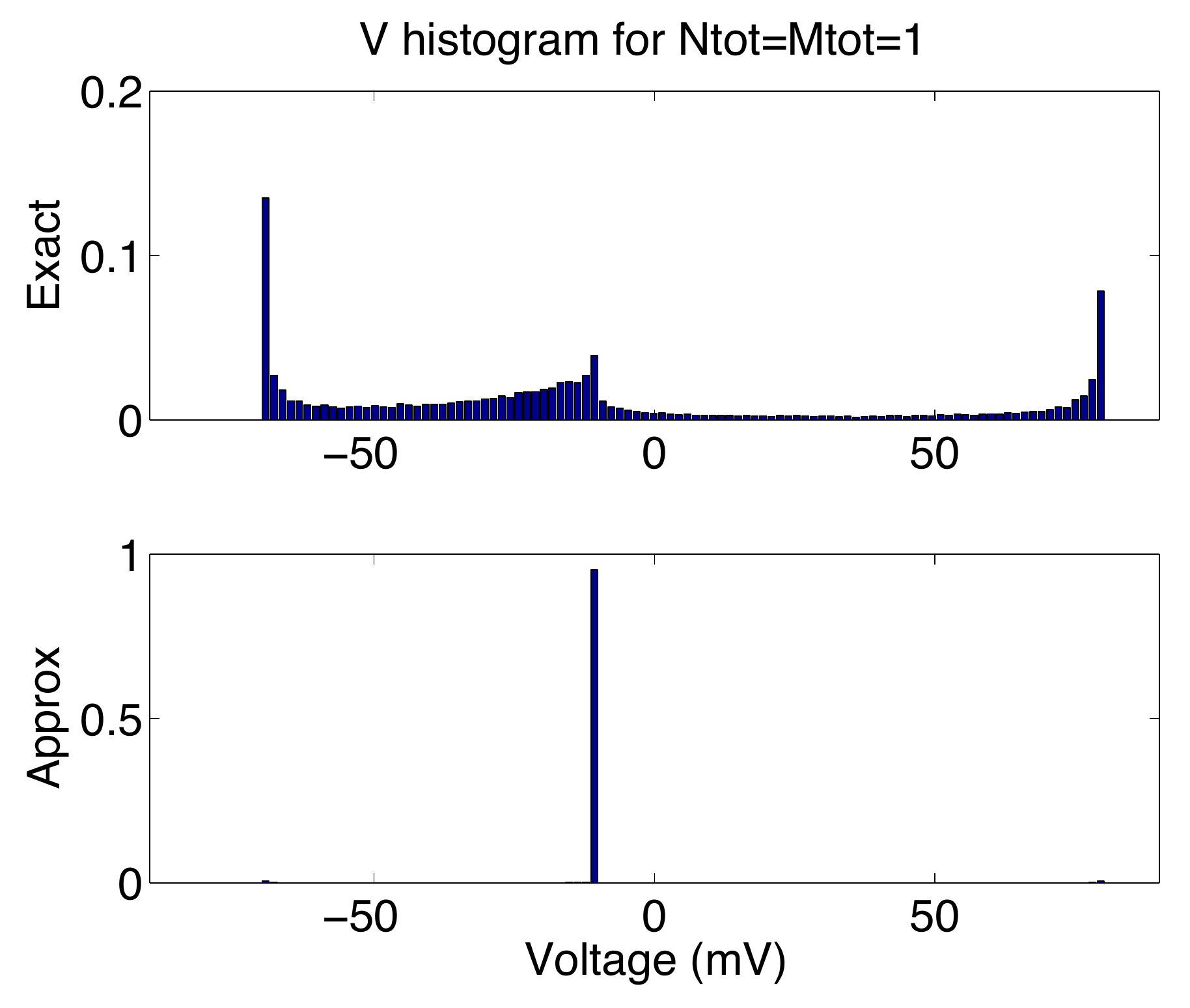} 
   \includegraphics[width=3.2in]{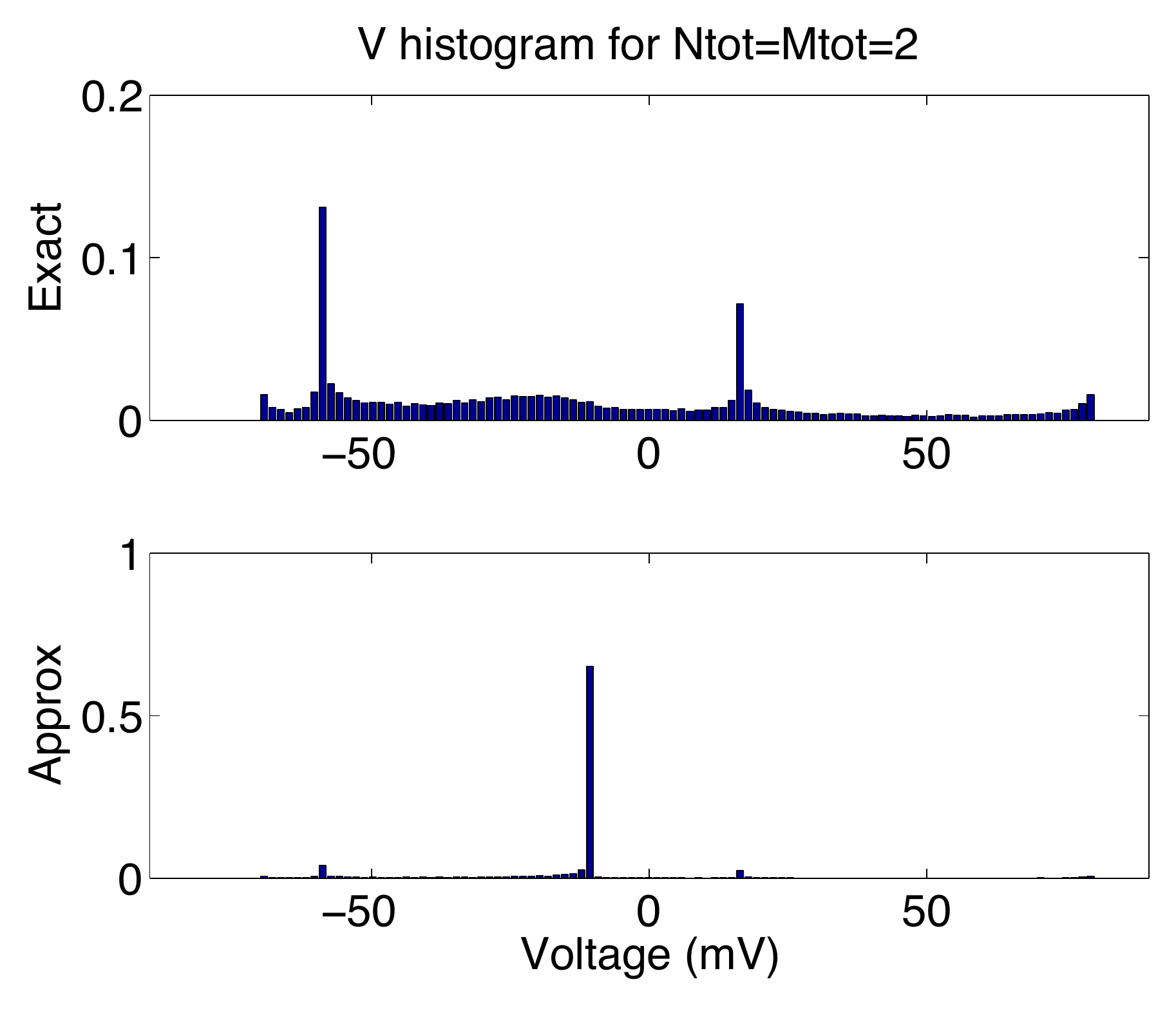} 
   \includegraphics[width=3.2in]{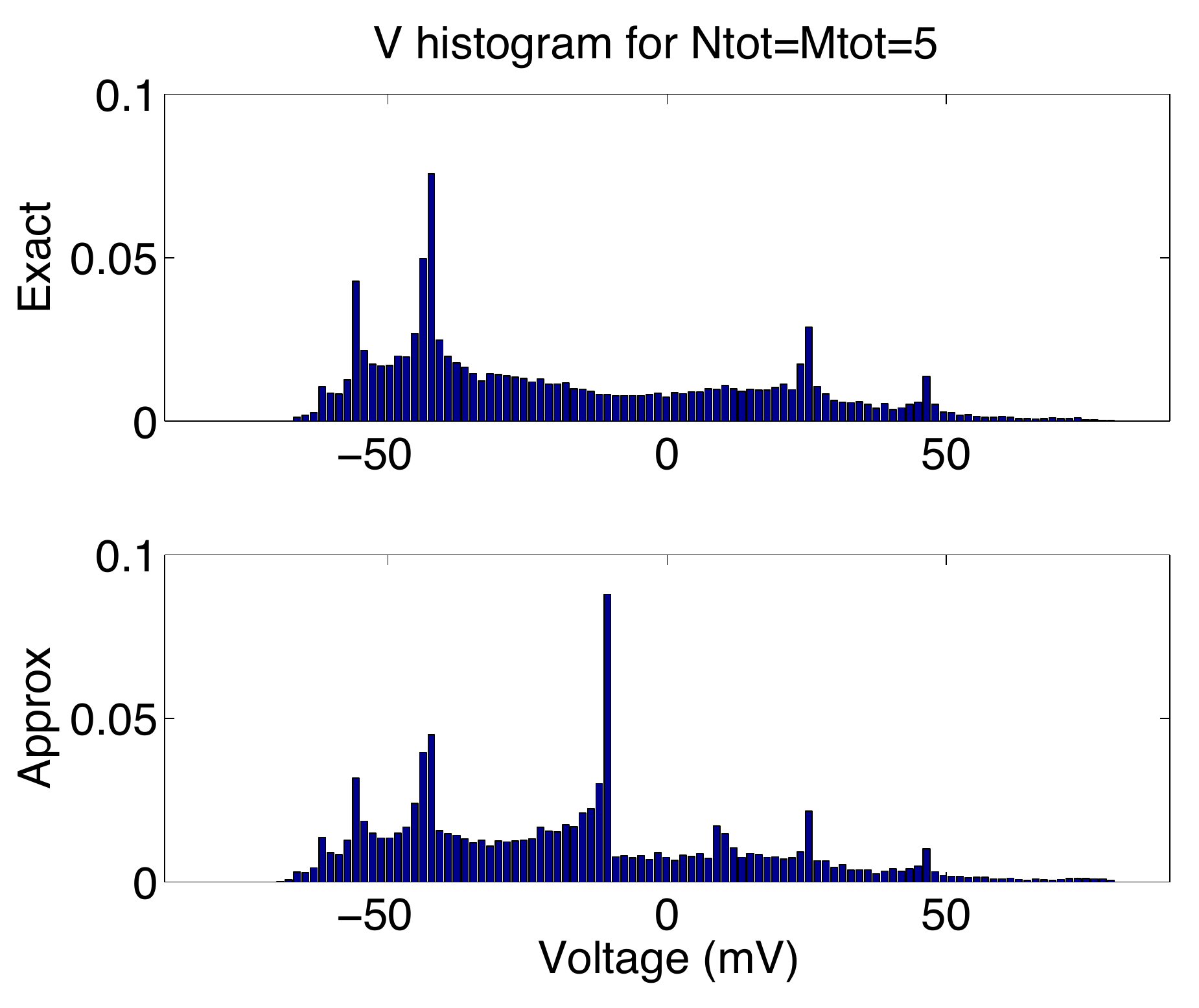} 
   \includegraphics[width=3.2in]{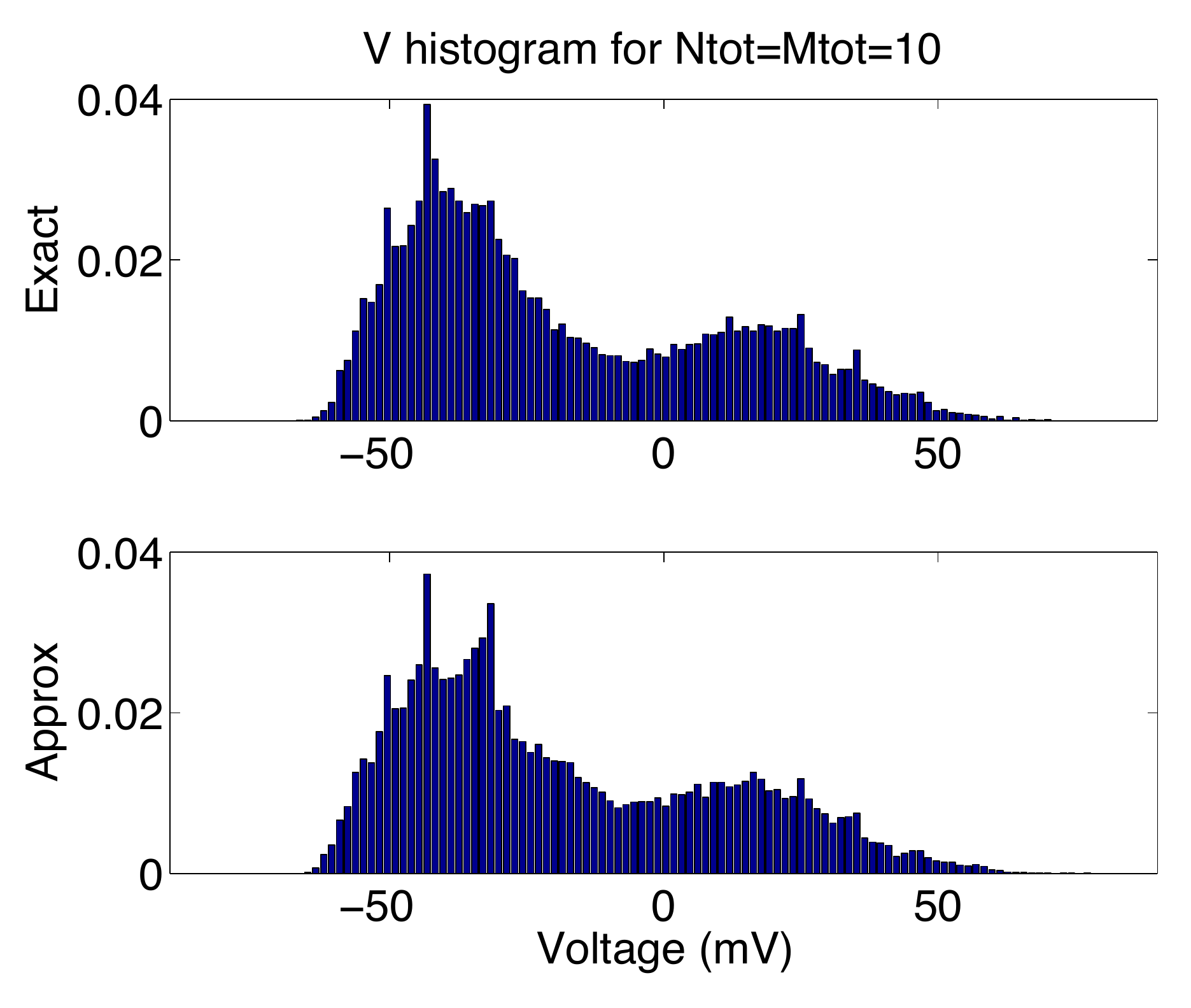} 
   \includegraphics[width=3.2in]{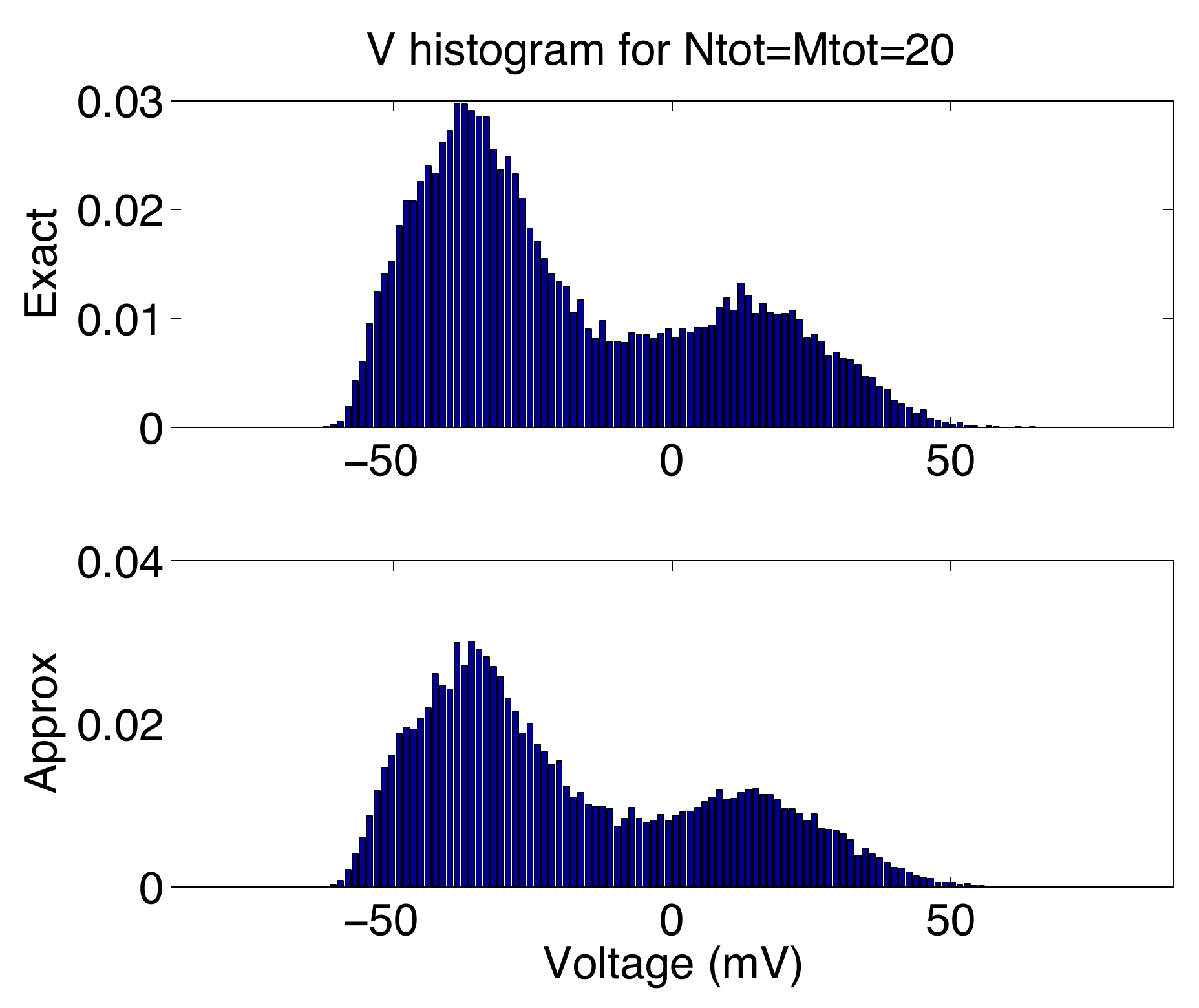} 
   \includegraphics[width=3.2in]{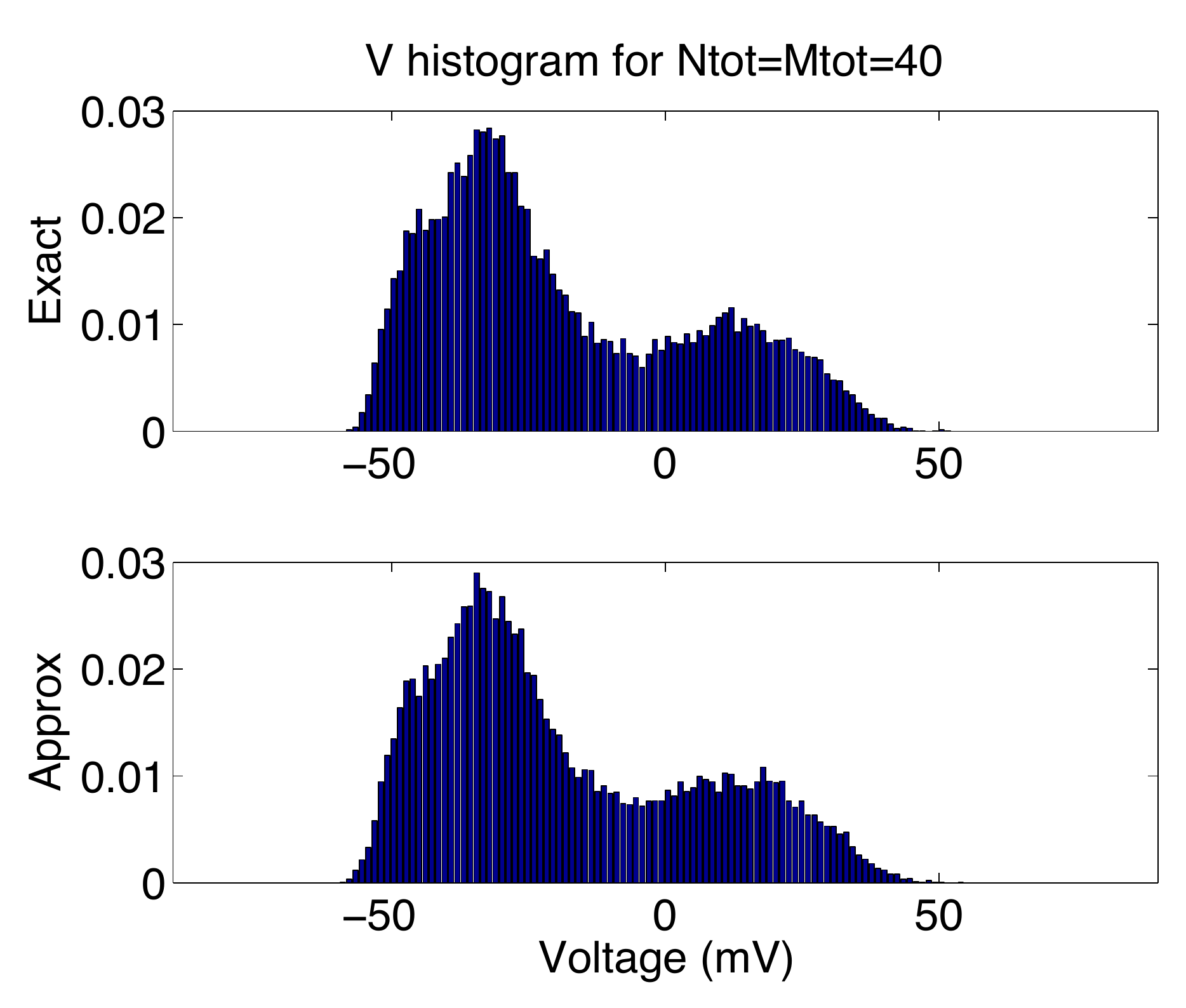} 
   \caption{Histograms of voltage for exact and \bl{approximate piecewise constant} algorithms. The $V$-axis was partitioned into 100 equal width bins for each pair of histograms. }
   \label{fig:v-histograms}
\end{figure}

To quantify the similarity of the histograms we calculated the empirical $L_1$ difference between the histograms in two ways: first for the full $(v,n,m)$ histograms, and then for the histograms collapsed  to the voltage axis. 
Let $\rho_{m,n}(v)$ and $\tilde{\rho}_{m,n}(v)$ denote the stationary distributions for the exact and the approximate algorithms, respectively (assuming these exist).  To compare the two distributions we approximate the $L_1$ distance between them, i.e.
\begin{equation*}
d(\rho,\tilde{\rho})=\int_{v_\text{min}}^{v_\text{max}}\left(\sum_{m=0}^{M_\text{tot}}\sum_{n=0}^{N_\text{tot}}|\rho_{m,n}(v)-\tilde{\rho}_{m,n}(v)|\right)\,dv
\end{equation*}
where $v_\text{min}$ and $v_\text{max}$ are chosen so that for all $m,n$ we have $F(v_\text{min},n,m)>0$ and $F(v_\text{max},n,m)<0$.  
{It is easy to see that such values of $v_\text{min}$ and $v_\text{max}$ must exist, since $F(v,n,m)$ is a linear and monotonically decreasing function of $v$ for any fixed pair $(n,m)$, cf.~equation \eqref{MLdet-v-full}.  Therefore, for any exact simulation algorithm,} once the voltage component of a trajectory falls in the interval $v_\text{min}\le v \le v_\text{max}$, it remains in this interval for all time. %{(Dave: Peter, where you going to reference your work and make clear that this is possible?)} 
We approximate the voltage integral by summing over 100 evenly spaced bins along the voltage axis. 
Figure \ref{fig:histogram}, top panel, shows empirical estimates of $d(\rho,\tilde{\rho})$ for values of $M_{\text{tot}}=N_{\text{tot}}$ ranging from 1 to 40.
The bottom panel shows the histogram for voltage considered alone.}

\begin{figure}[htbp] %  figure placement: here, top, bottom, or page
   \centering
   \includegraphics[width=5in]{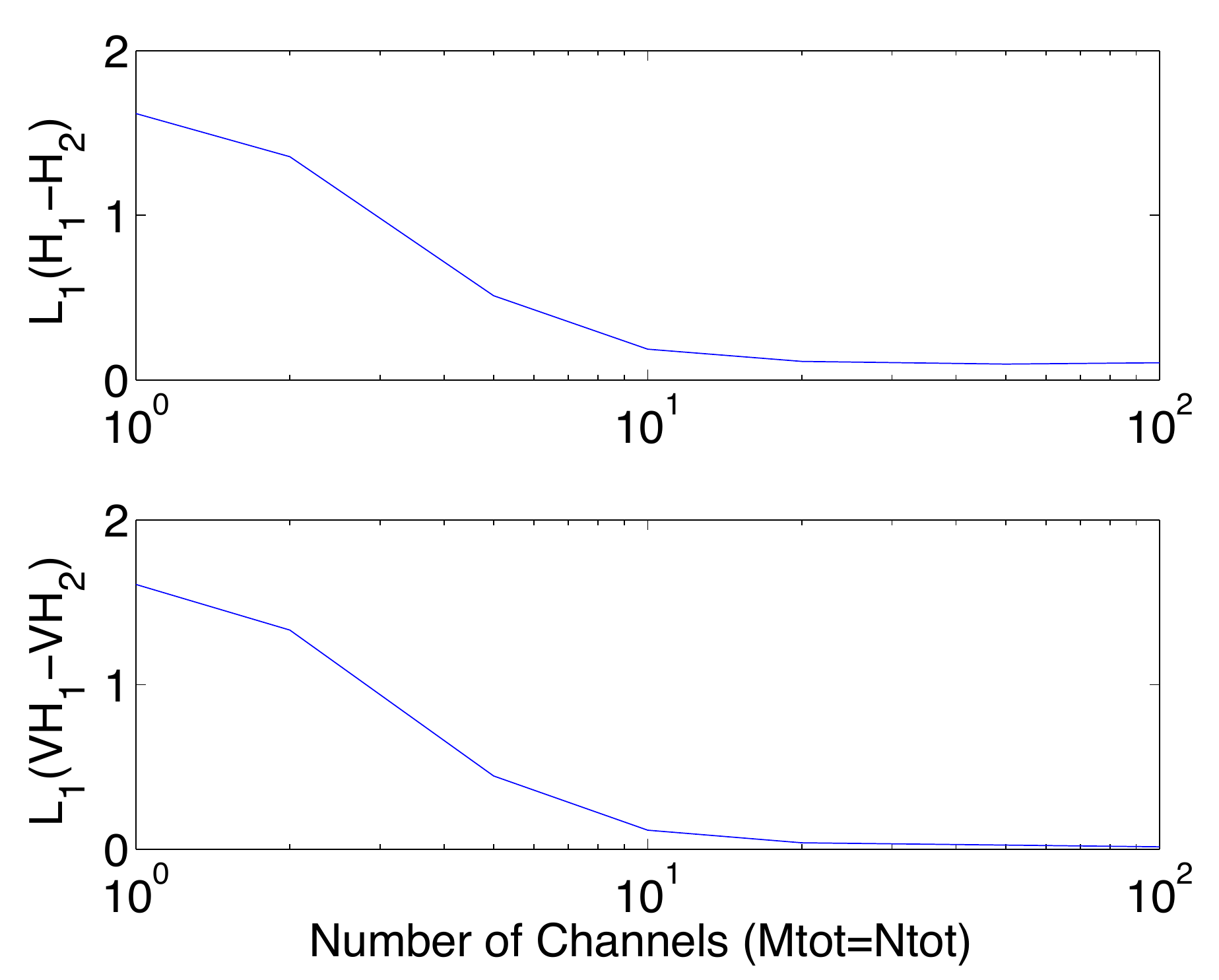} 
   \caption{$L_1$ differences between empirical histograms generated by exact and approximate algorithms. $H_j(v,n,m)$ is the  number of samples for which $V\in[v,v+\Delta v), N=n$ and $M=m$, where $\Delta v$ is a  discretization of the voltage axis into 100 equal width bins; $j=1$ represents an empirical histogram obtained by an exact algorithm (Algorithm \ref{main-algorithm}), and $j=2$ represents an empirical histogram obtained by a \bl{piecewise-constant propensity approximation}.  Note the normalized $L_1$ difference of two histograms can range from 0 to 2.  \bl{Based on 2,000,000 samples for each algorithm.}
   \textbf{Top:} Difference for the full histograms, $L_1(H_1-H_2)\equiv\sum_{v,n,m}|H_1(v,n,m)-H_2(v,n,m)|/\# \text{samples}$.  Here $v$ refers to binned voltage values. \bl{The total number of bins is $100\,M_\text{tot}N_\text{tot}$.}
   \textbf{Bottom:} Difference for the corresponding voltage histograms, $L_1(VH_1-VH_2)\equiv\sum_v\left|VH_1(v)-VH_2(v)\right|/\# \text{samples}$, where $VH_j(v)=\sum_{n,m}H_j(v,n,m)$.  \bl{The total number of bins is 100.} }
   \label{fig:histogram}
\end{figure}

\newpage

\section{Coupling, variance reduction, and parametric sensitivities}
\label{sec:coupling}
It is relatively straightforward to derive the exact simulation strategies from the different representations provided here.  Less obvious, and perhaps even more useful, is the fact that the random time change formalism also lends itself to the development of new computational methods that potentially allow for significantly greater computational speed,  with no loss in accuracy.  They do so by providing ways to \textit{couple} two processes in order to reduce the variance, and thereby increase the speed, of different natural estimators.  

%We show how this can be done for two common problems: (i)  the computation of expectations (including means, variances, probabilities, etc.) with multi-level Monte Carlo, and (ii) the computation of parametric sensitivities.

%\subsection{Variance reduction methods for parametric sensitivities} 

For an example of a common computational problem where such a benefit can be had, consider the computation of \textit{parametric sensitivities},  which is a powerful tool in determining parameters to which a system output is most responsive.  Specifically, suppose that the intensity/propensity functions are dependent on some vector of parameters $\theta$.  For instance, $\theta$ may represent a subset of the system's mass action kinetics constants, the cell capacitance, or the underlying number of channels of each type. 
We may wish to know how sensitively a quantity such as the average firing rate or interspike interval variance depends on the parameters $\theta$.
 We then consider a family of  models $\left(V^\theta,X^\theta\right)$, parameterized by $\theta$, with stochastic equations 
\begin{align}\label{eq:main_theta}
\begin{split}
\frac{d}{dt}V^\theta(t) &= f\left(\theta,V^\theta(t),X^\theta(t)\right)\\
X^{\theta}(t)&=X^\theta_0+\sum_{k} Y_k\left(\int_0^t
\lambda_k^\theta\left(V^\theta(s),X^\theta (s)\right)\, ds\right)\zeta_k,
\end{split}
\end{align}
where $f$ is some function, and all other notation is as before.
Some quantities arising in neural models depend on the entire sample path, such as the mean firing rate and the interspike interval variance.
Let $g\left(\theta,V^\theta,X^\theta \right)$ by a path functional capturing the quantity of interest.  In order to efficiently and accurately evaluate the relative shift in the expectation of $g$ due to a perturbation of the parameter vector, we would estimate 
\begin{eqnarray}\nonumber
	\tilde s &=& \epsilon^{-1}\E \left[ g\left(\theta', V^{\theta'},X^{\theta' }\right) -  g\left(\theta,V^\theta, X^{\theta}\right)\right] \\
	&\approx& \frac{1}{\epsilon N}\sum_{i = 1}^N  \left[ g\left(\theta',V_{[i]}^{\theta'}, X_{[i]}^{\theta'}\right) -  g\left(\theta, V_{[i]}^{\theta},X_{[i]}^{\theta}\right)\right]
	\label{eq:s}
\end{eqnarray}
where $\epsilon=||\theta-\theta' ||$ and where $\left(V_{[i]}^{\theta},X_{[i]}^{\theta}\right)$ represents the $i$th path generated with a parameter choice of $\theta$, and $N$ is the number of sample paths computed for the estimation.  That is, we would use a finite difference and Monte Carlo sampling to approximate the change in the expectation.  

 If the paths  $\left(V_{[i]}^{\theta'},X_{[i]}^{\theta'}\right)$ and $\left(V_{[i]}^{\theta},X_{[i]}^{\theta}\right)$ are generated independently, then the variance of the  estimator \eqref{eq:s} for $\tilde s$ is $O(N^{-1}\epsilon^{-2})$, and the standard deviation of the estimator is  $O(N^{-1/2} \epsilon^{-1})$.  This implies that in order to reduce the confidence interval of the estimator to a target  level of $\rho >0$, we require
 \[
 	N^{-1/2} \epsilon^{-1} \lesssim \rho \implies N \gtrsim \epsilon^{-2}\rho^{-2},
 \]
 which can be prohibitive.
 Reducing the variance of the estimator can be achieved by  \textit{coupling} the processes $\left(V_{[i]}^{\theta'},X_{[i]}^{\theta'}\right)$ and $\left(V_{[i]}^\theta,X_{[i]}^{\theta}\right)$,  so that they are correlated, by constructing them on the same probability space.  
 %This process is analogous to the simulation paths according to the random time change algorithm and a piecewise constant propensity approximation algorithm used to illustrate strong divergence (Figure \ref{fig:compare}).   
 The representations we present here lead rather directly to schemes for reducing the variance of the estimator.  We discuss different possibilities.
 \begin{enumerate}
 \item  \textbf{The common random number (CRN) method.}   The CRN method simulates both processes according to the Gillespie representation \eqref{eq:Gill_when}-\eqref{eq:Voltage2} with the same Poisson process $Y$ and the same stream of uniform random variables $\{\xi_i\}$.  In terms of implementation, one simply reuses the same two streams of uniform random variables in Algorithm \ref{gill-algorithm} for both processes.  
 \item \textbf{The common reaction path method (CRP).} The CRP method simulates both processes according to the random time change representation with the same Poisson processes $Y_k$.   In terms of implementation, one simply makes one stream of uniform random variables \textit{for each reaction channel}, and then uses these streams for the simulation of both processes.  See \cite{Khammash2010}, where this method was introduced in the context of biochemical models.

 \item  \textbf{The coupled finite difference method (CFD).}  The CFD method utilizes a  ``split coupling''  introduced in \cite{AndCFD2012}.  Specifically, it splits the  counting process for each of the reaction channels into three pieces: one counting process  that is shared by $X^\theta$ and $X^{\theta'}$  (and has propensity equal to the minimum of their respective intensities for that reaction channel), one that only accounts for the jumps of $X^\theta$, and one that only accounts for the jumps of $X^{\theta'}$.  Letting $a\wedge b = \min\{a,b\}$, the precise coupling is  
 \begin{align}
\begin{split}
	X^{\theta'}(t) &=  X_0^{\theta'} + \sum_{k} Y_{k,1}\left( \int_0^t m_k(\theta,\theta',s)  \, ds  \right)\zeta_k\\
	&+ \sum_{k} Y_{k,2}\left(  \int_0^t \lambda_k^{\theta'}(V^{\theta'}(s),X^{\theta'}(s))    -  m_k(\theta,\theta',s)  \, ds \right)\zeta_k\\
	X^{\theta}(t) &=  X^{\theta}(t) + \sum_{k} Y_{k,1}\left( \int_0^t m_k(\theta,\theta',s) \, ds  \right)\zeta_k\\
	&+ \sum_{k} Y_{k,3}\left(\int_0^t   \lambda_k^{\theta}(V^\theta(s),X^{\theta}(s))  -  m_k(\theta,\theta',s) \, ds  \right)\zeta_k,
    \end{split}
    \label{eq:CFD}
\end{align}
where
\[
	m_k(\theta,\theta',s) \equiv \lambda_k^\theta(V^\theta(s),X^\theta(s)) \wedge \lambda_k^{\theta'}(V^{\theta'}(s),X^{\theta'}(s)),
\]
and where $\{Y_{k,1},Y_{k,2},Y_{k,3}\}$ are independent unit-rate Poisson processes.  Implementation is then carried out by Algorithm \ref{main-algorithm} in the obvious manner.
\end{enumerate}

While the different representations provided in this paper imply different exact simulation strategies (i.e.~Algorithms \ref{main-algorithm} and \ref{gill-algorithm}), those strategies still produce statistically equivalent paths.  This is \textit{not} the case for the methods for parametric sensitivities provided above.  To be precise, each of the methods  constructs a coupled pair of processes $\left((V^\theta,X^\theta),(V^{\theta'},X^{\theta'})\right)$, and the marginal processes $\left(V^\theta,X^\theta\right)$ and $\left(V^{\theta'},X^{\theta'}\right)$ are all statistically equivalent no matter the method used.  However, the covariance $\textsf{Cov}\left((V^\theta(t),X^\theta(t)),(V^{\theta'}(t),X^{\theta'}(t))\right)$ can be drastically different.  This is important, since it is variance reduction we are after, and for any component $X_j$,
\[
	\textsf{Var}(X_j^\theta(t) - X_j^{\theta'}(t)) = \textsf{Var}(X_j^\theta(t)) + \textsf{Var}(X_j^{\theta'}(t)) - 2\textsf{Cov}(X_j^\theta(t),X_j^{\theta'}).
\]
Thus, minimizing variance is equivalent to maximizing covariance.  Typically, the CRN method does the worst job of maximizing the covariance, even though it is the most widely used method \cite{rishi2013}.  The CFD method with the split coupling procedure typically does the best job of maximizing the covariance, though examples exist in which the CRP method is the most efficient \cite{AndCFD2012,AK2014,rishi2013}.

\section{Discussion}
\label{sec:future}

We have provided two general representations for stochastic ion channel kinetics, {one based on the random time change formalism, and one extending Gillespie's algorithm to the case of ion channels driven by time-varying membrane potential.}  \bl{These representations are  known in other branches of science, but less so in the neuroscience community.}  We believe that the {random time change representation (Algorithm \ref{main-algorithm},  \eqref{eq:RTC_main}-\eqref{eq:voltage})} will be particularly useful to the {computational neuroscience} community as it allows for generalizations of computational methods developed in the context of biochemistry, in which the propensities depend upon the state of the jump process only.   For example, variance reduction strategies for the efficient computation of first and second order sensitivities  \cite{AndCFD2012,AndSkubak,Khammash2010}, as discussed in Section \ref{sec:coupling}, and for the efficient computation of expectations using multi-level Monte Carlo \cite{AndHigham2012, Giles2008} now become feasible.

The random time change approach avoids several  approximations that are commonly found in the literature.  In simulation algorithms based on a fixed time step chemical Langevin approach, it is necessary to assume that the increments in channel state are approximately Gaussian distributed over an appropriate time interval \cite{FoxLu1994PRE,GoldwynImennovFamulareShea-Brown2011PRE,GoldwynShea-Brown2011PLoSCB,MinoRubinsteinWhite2002AnnBiomedEng,White+Chow+Ritt+Soto-Trevino+Kopell:1998:JCN}.  However, in exact simulations with small membrane patches corresponding to $M_\text{tot}=40$ calcium channels, the exact algorithm routinely visits the states $M(t)=0$ and $M(t)=40$, for which the Gaussian increment approximation {is} invalid regardless of time step size.  The main alternative algorithm typically found in the literature is the piecewise constant propensity or approximate forward algorithm \cite{FischSchwalgerLindnerHerzBenda2012JNSci,KisperskyTilman2008Scholarpedia,SchwalgerFischBendaLindner2010PLosCB}.  {However, this algorithm ignores changes to membrane potential during the intervals between channel state changes.  As the sensitivity of ion channel opening and closing to voltage is fundamental to the neurophysiology of cellular excitability, these algorithms are not appropriate unless the time between openings and closing is especially small.}
% The sensitivity of ion channel opening and closing to voltage is fundamental to the neurophysiology of cellular excitability, but changes to membrane potential {are} ignored during the intervals  between channel state changes in the approximate forward algorithm.  
The exact algorithm \cite{ClayDeFelice1983BiophysJ,KeenerNewby2011PRE,NewbyBressloffKeener2013PRL} is  straightforward to implement and avoids these approximations {and pitfalls}.

\bl{Our study restricts attention to the suprathreshold regime of the Morris-Lecar model, in which the applied current (here, $I_\text{app}=100$) puts the deterministic system well above the Hopf bifurcation marking the onset of oscillations.  In this regime, spiking is not the result of noise-facilitated release, as it is in the subthreshold or excitable regime.  Bressloff, Keener and Newby used eigenfunction expansion methods, path integrals, and the theory of large deviations to study spike initiation as a noise facilitated escape problem in the excitable regime, as well as to incorporate synaptic currents into a stochastic network model; they use versions of the exact algorithm presented in this paper \cite{BressloffNewby2014PRE,NewbyBressloffKeener2013PRL}.
Interestingly, these authors find that the usual separation-of-time scales picture breaks down. That is, the firing rate obtained by 1D Kramers rate theory when the slow recovery variable (fraction of open potassium gates) is taken to be fixed does not match that obtained by direct simulation with the exact algorithm.  Moreover, by considering a maximum likelihood approach to the 2D escape problem, the authors show that, counterintuitively, spontaneous closure of potassium channels contributes more significantly to noise induced escape than spontaneous opening of sodium channels, as  might naively have been expected.
}

\bl{The algorithms presented here are broadly applicable beyond the effects of channel noise on the regularity of action potential firing in a single compartment neuron model.  Exact simulation of hybrid stochastic models have been used, for instance, to study spontaneous dendritic NMDA spikes \cite{BressloffNewby2014PhysBiol},  intracellular growth of the T7 bacteriophage \cite{AlfonsiCancesTuriniciVenturaHuisinga2005ESIAM}, and hybrid stochastic network models taking into account piecewise deterministic synaptic currents \cite{BressloffNewby2013SIADS}.  This latter work represents a significant extension of the neural master equation approach to stochastic population models \cite{Bressloff20009SIAM,BuiceCowanChow2010NeCo}.}

{For any simulation algorithm, it is reasonable to ask about the growth of complexity of the algorithm as the underlying stochastic model is enriched.  For example, the natural jump Markov interpretation of Hodgkin and Huxley's model for the sodium channel comprises eight distinct states with twenty different state-to-state transitions, each driven by an independent Poisson process in the random time change representation.  Recent investigations of sodium channel kinetics have led neurophysiologists to formulate models with as many as 26 distinct states connected by 72 transitions \cite{MilescuYamanishiPtakSmith2010JNsci}.  While the random time change representation extends naturally to such scenarios, it may also be fruitful to combine it with complexity reduction methods such as the stochastic shielding algorithm introduced by Schmandt and Gal\'{a}n \cite{SchmandtGalan2012PRE}, and analyzed by Schmidt and Thomas \cite{SchmidtThomas2014JMN}.  For example, of the twenty independent Poisson processes driving a discrete Markov model of the Hodgkin-Huxley sodium channel, only four of the processes directly affect the conductance of the channel; fluctuations associated with the remaining sixteen Poisson processes may be ignored with negligible loss of accuracy in the first and second moments  of the channel occupancy.  Similarly, for the 26 state sodium channel model proposed in \cite{MilescuYamanishiPtakSmith2010JNsci}, all but 12 of the 72 Poisson processes representing state-to-state transitions can be replaced by their expected values.  Analysis of algorithms combining stochastic shielding and the random time change framework are a promising direction for future research.} 

\vspace{.2in}
\noindent\textbf{Acknowledgments.}
	
	Anderson was supported by NSF grant  DMS-1318832  and Army Research Office grant W911NF-14-1-0401. Ermentrout was supported by NSF grant DMS-1219754.  
	Thomas was supported by NSF grants EF-1038677, DMS-1010434, and DMS-1413770, by a grant from the Simons Foundation (\#259837), and by the Council for the International Exchange of Scholars (CIES).    	
	  We gratefully acknowledge the Mathematical Biosciences Institute (MBI, supported by NSF grant DMS 0931642) at The Ohio State University for hosting a workshop at which this research was initiated.   
	The authors thank David Friel and Casey Bennett for helpful discussions and testing of the algorithms.

\newpage
\appendix
\section{Sample Implementations of Random Time Change Algorithm (Algorithm 1)}
\label{app:code}
%{PJT: I still have to go through the sample code to make sure it is presented OK.}

\subsection{Morris Lecar with Stochastic Potassium Channel}

\subsubsection{XPP: ml-rtc-konly.ode}
\label{sec:xpp-k-only}
\begin{verbatim}
# ML with stochastic potassium channels
# uses the exact random time change algorithm

# membrane potential
v'=(I-gca*minf*(V-Vca)-gk*wtot*(V-VK)-gl*(V-Vl))/c
# unit exponentials
t[3..4]'=0
# number of potassium(w) /calcium (m) channels open
w'=0
# int_0^t beta(V(s)) ds 
# for the 4 reactions
awp'=aw
bwp'=bw
# initialize unit exponentials
t[3..4](0)=-log(ran(1))
# look for crossings, reset integrals, increment channels, choose next time
global 1 awp-t3 {t3=-log(ran(1));awp=0;w=w+1}
global 1 bwp-t4 {t4=-log(ran(1));bwp=0;w=w-1}
# parameters
par Nw=100
init v=-50
# fraction of open channels
wtot=w/Nw
# ML channel kinetic definitions
minf=.5*(1+tanh((v-va)/vb))
winf=.5*(1+tanh((v-vc)/vd))
tauw=1/cosh((v-vc)/(2*vd))
alw=winf/tauw
blw=1/tauw-alw
# independent, so rates are just multiples of number in each state
aw=alw*(Nw-w)*phi
bw=blw*w*phi
param vk=-84,vl=-60,vca=120
param i=75,gk=8,gl=2,c=20,phim=.4
param va=-1.2,vb=18
param vc=2,vd=30,phi=.04,gca=4.4
# set up some numerics and plotting stuff
@ dt=.01,nout=100,total=4000,bound=1000
@ maxstor=5000,meth=euler
@ xhi=4000,ylo=-65,yhi=50
done
\end{verbatim}

\newpage
\subsubsection{Matlab: ml\_rtc\_exact\_konly.m}
\label{sec:matlab-k-only}
\begin{verbatim}
function [V,N,t,Ntot]=ml_rtc_exact_konly(tmax,Ntot)

%function [V,N,t,Ntot]=ml_rtc_exact_konly(tmax,Ntot);
%
% Exact solution of Morris Lecar with stochastic potassium channel. 
% Using the random time change algorithm.  We track two reactions: 
%
% Rxn 1: closed -> open (per capita rate alpha)
% Rxn 2: open -> closed (per capita rate beta)
%
% Default Ntot=40, tmax=4000.  
%
% Author: PJT July 2013, Case Western Reserve University.

% Applied current "Iapp" set on line 47 below.

%% Use global variables to represent the channel state 
%    and random trigger for Poisson process
global Npotassium_tot   % total number of potassium channels
global Npotassium       % number of open potassium channels 
global tau1  T1         % time to next opening event (internal to reaction 1)
global tau2  T2         % time to next closing event (internal to reaction 2)

%% Set defaults for input arguments
if nargin < 2, Ntot=40; end
Npotassium_tot=Ntot;
if nargin < 1, tmax=4e3; end
%% Parameters
% Standard Morris-Lecar parameters giving a globally attracting limit cycle
% (if applied current Iapp=100) or a stable fixed point (if Iapp=75).
va = -1.2;
vb=18;
vc = 2;
vd = 30;
phi = 0.04;
%% Functions for Morris-Lecar
global Iapp; Iapp=@(t)100; % applied current. 
global minf; minf=@(v)0.5*(1+tanh((v-va)/vb)); % m-gate activation 
global xi; xi=@(v)(v-vc)/vd;  % scaled argument for n-gate input
global ninf; ninf=@(v)0.5*(1+tanh(xi(v)));  % n-gate activation function
global tau_n; tau_n=@(v)1./(phi*cosh(xi(v)/2)); % n-gate activation t-const
global alpha; alpha=@(v)(ninf(v)./tau_n(v)); % per capita opening rate
global beta; beta=@(v)((1-ninf(v))./tau_n(v)); % per capita closing rate
%% ODE options including reset
options=odeset('Events',@nextevent);
%% Initialize
t0=0;
t=t0; % global "external" time
tau1=-log(rand); % time of next event on reaction stream 1 ("internal time")
tau2=-log(rand); % time of next event on reaction stream 2 ("internal time")
T1=0; % integrated intensity function for reaction 1
T2=0; % integrated intensity function for reaction 2
V0=-50; % start at an arbitrary middle voltage
V=V0;
N0=ceil(Ntot/2); % start with half of channels open
Npotassium=N0;
N=N0; % use N to record the time course
%% Loop over events
while t(end)<tmax
%% Integrate ODE for voltage, until the next event is triggered
% State vector U for integration contains the following components
%     [voltage; 
%      integral of (# closed)*alpha(v(t)); 
%      integral of (# open)*beta(v(t));
%      # open].
U0=[V0;0;0;N0];
tspan=[t(end),tmax];
[tout,Uout,~,~,event_idx]=ode23(@dudtfunc,tspan,U0,options);
Vout=Uout(:,1); % voltage at time of next event
Nout=Uout(:,4); % number of channels open at end of next event
t=[t,tout'];
V=[V,Vout'];
N=[N,Nout'];
%% Identify which reaction occurred, adjust state, and 
%  continue, and set trigger for next event.
mu=event_idx; % next reaction index
if mu==1 % next reaction is a channel opening
    N0=N0+1;    % increment channel state
    tau1=tau1-log(rand); % increment in tau1 is exp'l with mean 1
elseif mu==2 % next reaction is another channel closing
    N0=N0-1;    % decrement channel state
    tau2=tau2-log(rand); % increment in tau2 likewise
end
Npotassium=N0;
if N0>Npotassium_tot, error('N>Ntot'), end
if N0<0, error('N<0'), end
T1=T1+Uout(end,2);
T2=T2+Uout(end,3);
V0=V(end);
end % while t(end)<tmax
%% Plot output
figure
subplot(4,1,1),plot(t,N),xlabel('time'),ylabel('N')
subplot(4,1,2:3),plot(V,N,'.-'),xlabel('V'),ylabel('N')
subplot(4,1,4),plot(t,V),xlabel('time'),ylabel('V')
shg
end % End of function ml_rtc_exact_konly
%% Define the RHS for Morris-Lecar
% state vector for integration is as follows:
% u(1) = voltage
% u(2) = integral of activation hazard function
% u(3) = integral of inactivation hazard function
% u(4) = N (number of open potassium channels)
function dudt=dudtfunc(t,u)
global Iapp  % Applied Current
global minf  % asymptotic target for (deterministic) calcium channel
global Npotassium Npotassium_tot  % num. open, total num. of channels
global alpha beta  % per capita transition rates
%% Parameters
vK = -84; vL = -60; vCa = 120; 
gK =8; gL =2; C=20; gCa = 4.4;
%% calculate the RHS;
v=u(1); % extract the voltage from the input vector
dudt=[(Iapp(t)-gCa*minf(v)*(v-vCa)-gL*(v-vL)-...
        gK*(Npotassium/Npotassium_tot)*(v-vK))/C; % voltage
    alpha(v)*(Npotassium_tot-Npotassium); % channel opening internal time
    beta(v)*Npotassium;% channel closing internal time 
    0]; % N is constant between events
end
%% Define behavior at threshold crossing
function [value,isterminal,direction] = nextevent(~,u)
    global tau1 T1 % timing trigger for reaction 1 (opening)
    global tau2 T2 % timing trigger for reaction 2 (closing)
    value=[u(2)-(tau1-T1);u(3)-(tau2-T2)]; 
    isterminal=[1;1]; % stop and restart integration at crossing
    direction=[1;1]; % increasing value of the quantity at the trigger 
end
\end{verbatim}

\newpage
\subsection{Morris Lecar with Stochastic Potassium and Calcium Channels}

\subsubsection{XPP: ml-rtc-exact.ode}
\label{sec:xpp-both}
\begin{verbatim}
# ML with both potassium and calcium channels stochastic
# uses the exact random time change algorithm.

# membrane potential
v'=(I-gca*mtot*(V-Vca)-gk*wtot*(V-VK)-gl*(V-Vl))/c

# unit exponentials
t[1..4]'=0

# number of potassium(w) /calcium (m) channels open
w'=0
m'=0

# int_0^t beta(V(s)) ds 
# for the 4 reactions
amp'=am
bmp'=bm
awp'=aw
bwp'=bw

# initialize unit exponentials
t[1..4](0)=-log(ran(1))

# look for crossings, reset integrals, increment channels, choose next time
global 1 amp-t1 {t1=-log(ran(1));amp=0;m=m+1}
global 1 bmp-t2 {t2=-log(ran(1));bmp=0;m=m-1}
global 1 awp-t3 {t3=-log(ran(1));awp=0;w=w+1}
global 1 bwp-t4 {t4=-log(ran(1));bwp=0;w=w-1}

# parameters
par Nm=100,Nw=100
init v=-50
param vk=-84,vl=-60,vca=120
param i=75,gk=8,gl=2,c=20,phim=.4
param va=-1.2,vb=18
param vc=2,vd=30,phi=.04,gca=4.4

# fraction of open channels
wtot=w/Nw
mtot=m/Nm

# ML channel kinetic definitions
minf=.5*(1+tanh((v-va)/vb))
winf=.5*(1+tanh((v-vc)/vd))
tauw=1/cosh((v-vc)/(2*vd))
taum=1/cosh((v-va)/(2*vb))
alm=minf/taum
blm=1/taum-alm
alw=winf/tauw
blw=1/tauw-alw

# independent, so rates are just multiples of number in each state
am=alm*(Nm-m)*phim
bm=blm*m*phim
aw=alw*(Nw-w)*phi
bw=blw*w*phi

# set up some numerics and plotting stuff
@ dt=.01,nout=100,total=4000,bound=1000
@ maxstor=5000,meth=euler
@ xhi=4000,ylo=-65,yhi=50
done
\end{verbatim}

\newpage
\subsubsection{Matlab: mlexactboth.m}
\label{sec:matlab-both}
\begin{verbatim}
function [V,M,N,t,Mtot,Ntot]=mlexactboth(tmax,Mtot,Ntot)

%function [V,M,Mtot,N,Ntot]=mlexactboth(tmax,Mtot,Ntot);
%
% Exact solution of Morris Lecar with discrete stochastic potassium channel 
% (0<=N<=Ntot) and discrete stochastic calcium channel (0<=M<=Mtot).  
% Using the random time change representation, we track four reactions: 
%
% Rxn 1: calcium (m-gate) closed -> open (per capita rate alpha_m)
% Rxn 2: calcium (m-gate) open -> closed (per capita rate beta_m)
% Rxn 3: potassium (n-gate) closed -> open (per capita rate alpha_n)
% Rxn 4: potassium (n-gate) open -> closed (per capita rate beta_n)
%
% Default Mtot=40, Ntot=40, tmax=4000.
%
% Applied current "Iapp" set internally. 
%
% PJT June 2013, CWRU.  Following Bard Ermentrout's "ml-rtc-exact.ode".

%% Use global variables to represent the channel state 
% and random trigger for Poisson process

% Calcium
global Mcalcium_tot	% total number of calcium channels
global Mcalcium	% number of calcium channels in conducting state
global tau1  T1	% dummy variables for time to next Ca-opening event 
global tau2  T2	% dummy variables for time to next Ca-closing event 

% Potassium
global Npotassium_tot	% total number of potassium channels
global Npotassium	% number of potassium channels in conducting state 
global tau3 T3		% dummy variables for time to next K-opening event 
global tau4 T4		% dummy variables for time to next K-closing event 

%% Set defaults for input arguments

if nargin < 3, Ntot=40; end
Npotassium_tot=Ntot;
if nargin < 2, Mtot=40; end
Mcalcium_tot=Mtot;
if nargin < 1, tmax=4e3; end

%% Parameters
phi_m=0.4; 
va = -1.2; vb=18;
vc = 2; vd = 30; 
phi_n = 0.04;

%% Functions for Morris-Lecar
global Iapp; Iapp=@(t)100;		% applied current

global xi_m; xi_m=@(v)(v-va)/vb;	% scaled argument for m-gate input voltage
global minf; minf=@(v)0.5*(1+tanh(xi_m(v)));	% m-gate activation function
global tau_m; tau_m=@(v)1./(phi_m*cosh(xi_m(v)/2));	% m-gate time constant
global alpha_m; alpha_m=@(v)(minf(v)./tau_m(v));
global beta_m; beta_m=@(v)((1-minf(v))./tau_m(v));

global xi_n; xi_n=@(v)(v-vc)/vd;	% scaled argument for n-gate input
global ninf; ninf=@(v)0.5*(1+tanh(xi_n(v)));	% n-gate activation function
global tau_n; tau_n=@(v)1./(phi_n*cosh(xi_n(v)/2)); 	% n-gate time constant
global alpha_n; alpha_n=@(v)(ninf(v)./tau_n(v));
global beta_n; beta_n=@(v)((1-ninf(v))./tau_n(v));

%% ODE options including reset
options=odeset('Events',@nextevent);

%% initialize
t0=0;
t=t0; % global "external" time
tau1=-log(rand); % time of next event on reaction stream 1 ("internal time")
tau2=-log(rand); % time of next event on reaction stream 2 ("internal time")
tau3=-log(rand); % time of next event on reaction stream 3 ("internal time")
tau4=-log(rand); % time of next event on reaction stream 4 ("internal time")
T1=0; % integrated intensity function for reaction 1
T2=0; % integrated intensity function for reaction 2
T3=0; % integrated intensity function for reaction 3
T4=0; % integrated intensity function for reaction 4
V0=-50; % start at an arbitrary middle voltage
V=V0; % initial voltage
M0=0; % start with calcium channels all closed
Mcalcium=M0; % initial state of calcium channel
M=M0; % use M to record the time course

N0=ceil(Ntot/2); % start with half potassium channels open
Npotassium=N0; % initial state of potassium channel
N=N0; % use N to record the time course

%% Loop over events
while t(end)<tmax

%% integrate ODE for voltage, until the next event is triggered
% State vector U for integration contains the following components
%  1    [voltage; 
%  2    integral of (# closed)*alpha_m(v(t)); 
%  3    integral of (# open)*beta_m(v(t));
%  4    integral of (# closed)*alpha_n(v(t)); 
%  5    integral of (# open)*beta_n(v(t));
%  6    # open calcium channels;
%  7    # open potassium channels].
U0=[V0;0;0;0;0;M0;N0];
tspan=[t(end),tmax];
[tout,Uout,~,~,event_idx]=ode23(@dudtfunc,tspan,U0,options);
Vout=Uout(:,1); % voltage at time of next event
Mout=Uout(:,6); % number of calcium channels open at end of next event
Nout=Uout(:,7); % number of potassium channels open at end of next event
t=[t,tout'];
V=[V,Vout'];
M=[M,Mout'];
N=[N,Nout'];

%% Identify which reaction occurred, adjust state, and continue;
% and set trigger for next event.
mu=event_idx; % next reaction index
if mu==1 % next reaction is a calcium channel opening
    M0=M0+1; % increment calcium channel state
    tau1=tau1-log(rand); % increment in tau1 is exponentially distributed with mean 1
elseif mu==2 % next reaction is a calcium channel closing
    M0=M0-1; % decrement calcium channel state
    tau2=tau2-log(rand); % increment in tau2 is exponentially distributed with mean 1
elseif mu==3 % next reaction is a potassium channel opening
    N0=N0+1;    % increment potassium channel state
    tau3=tau3-log(rand); % increment in tau3 is exponentially distributed with mean 1
elseif mu==4 % next reaction is a potassium channel closing
    N0=N0-1;    % decrement potassium channel state
    tau4=tau4-log(rand); % increment in tau4 likewise
end
Mcalcium=M0;
Npotassium=N0;

if M0>Mcalcium_tot, error('M>Mtot'), end
if M0<0, error('M<0'), end
if N0>Npotassium_tot, error('N>Ntot'), end
if N0<0, error('N<0'), end

T1=T1+Uout(end,2);
T2=T2+Uout(end,3);
T3=T3+Uout(end,4);
T4=T4+Uout(end,5);
V0=V(end);

end % while t(end)<tmax

%% Plot output

figure
subplot(6,1,1),plot(t,M),ylabel('M','FontSize',20),set(gca,'FontSize',20)
subplot(6,1,2),plot(t,N),ylabel('N','FontSize',20),set(gca,'FontSize',20)
subplot(6,1,6),plot(t,V),xlabel('Time','FontSize',20)
    ylabel('V','FontSize',20),set(gca,'FontSize',20)
subplot(6,1,4:6),plot3(V,M,N,'.-'),xlabel('V','FontSize',20)
    ylabel('M','FontSize',20),zlabel('N','FontSize',20),set(gca,'FontSize',20)
grid on, rotate3d, shg

end % End of function mlexactboth

%% Define the RHS for Morris-Lecar
% state vector for integration is as follows:
% u(1) = voltage
% u(2) = integral of Ca-opening hazard function (Rxn 1)
% u(3) = integral of Ca-closing hazard function (Rxn 2)
% u(4) = integral of K-opening hazard function (Rxn 3)
% u(5) = integral of K-closing hazard function (Rxn 4)
% u(6) = M (number of open calcium channels)
% u(7) = N (number of open potassium channels)

function dudt=dudtfunc(t,u)
global Iapp % applied current
global Mcalcium Mcalcium_tot
global Npotassium Npotassium_tot
global alpha_m beta_m
global alpha_n beta_n

%% Parameters
vK = -84; vL = -60; vCa = 120; 
gK =8; gL =2; C=20; gCa = 4.4;

%% calculate the RHS;
v=u(1); % extract the voltage from the input vector
dudt=[...  % voltage
    (Iapp(t)-gCa*(Mcalcium/Mcalcium_tot)*(v-vCa)-gL*(v-vL)...
    -gK*(Npotassium/Npotassium_tot)*(v-vK))/C; 
    % Calcium chan. opening, internal time elapsed
    alpha_m(v)*(Mcalcium_tot-Mcalcium); 
    % Calcium chan. closing, internal time elapsed
    beta_m(v)*Mcalcium; 
    % Potassium chan. opening, internal time elapsed
    alpha_n(v)*(Npotassium_tot-Npotassium); 
    % Potassium chan. closing, internal time elapsed
    beta_n(v)*Npotassium; 
    0; % M is constant between events
    0]; % N is constant between events
end

%% define behavior at threshold crossing

function [value,isterminal,direction] = nextevent(~,u)
    global tau1 T1 % timing trigger for reaction 1 (calcium opening)
    global tau2 T2 % timing trigger for reaction 2 (calcium closing)
    global tau3 T3 % timing trigger for reaction 3 (potassium opening)
    global tau4 T4 % timing trigger for reaction 4 (potassium closing)
    value=[u(2)-(tau1-T1);u(3)-(tau2-T2);u(4)-(tau3-T3);u(5)-(tau4-T4)]; 
    isterminal=[1;1;1;1]; % stop and restart integration at crossing
    direction=[1;1;1;1]; % increasing value of the quantity at the trigger 
end

\end{verbatim}

%\textcolor{red}{DFA: what is the purpose of this appendix???}
%The deterministic Morris-Lecar equations \cite{ErmentroutTerman2010book,MorrisLecar1981BiophysJ,Rinzel+Ermentrout:1989}
%for voltage $v\in\mathbb{R}$ and potassium gate $n\in[0,1]$ represent a fast calcium current with gating variable $m$ as well as the slow potassium current:
%\begin{eqnarray}
%\frac{dv}{dt}&=&f(v,n)=\frac{1}{C}\left(I_{\text{app}}-g_{Ca}m_{\infty}(v)(v-v_{Ca}) - g_Kn(v-v_K)-g_L(v-v_L)\right)\\
%\frac{dn}{dt}&=&g(v,n)=\alpha(v)(1-n)+\beta(v)n
%=(n_{\infty}(v)-n)/\tau(v)
%\end{eqnarray}
%The terms $m_{\infty},\alpha,\beta,n_\infty$ and $\tau$ satisfy (for convenience we define $\xi=(v-v_c)/v_d$)
%\begin{eqnarray}
%m_{\infty}&=&\frac{1}{2}\left(1+\tanh\left(\frac{v-v_a}{v_b}\right)\right)\\
%\alpha(v)&=&\frac{\phi\cosh(\xi/2)}{1+e^{2\xi}}\\
%\beta(v)&=&\frac{\phi\cosh(\xi/2)}{1+e^{-2\xi}}\\
%n_{\infty}(v)&=&\left(1+\tanh\xi \right)/2\\
%\tau(v)&=&1/\left(\phi\cosh(\xi/2)\right).
%\end{eqnarray}
%We take standard values of the parameters (see \textit{e.g.}~Ermentrout's implementation -- see (\cite{ErmentroutTerman2010book}, \S 3)).  
%\begin{eqnarray}
%v_K&=&-84, v_L=-60,v_{Ca}=120\\
%I_{\text{app}}&=&100,g_K=8,g_L=2,C=20\\
%v_a&=&-1.2,v_b=18\\
%v_c&=&2,v_d=30,\phi=0.04,g_{Ca}=4.4
%\end{eqnarray}
%for which there is a globally attracting stable limit cycle.

\newpage
%\bibliographystyle{plain}
%\bibliography{math,neuroscience,PJT,Respiration,Dicty,reliability,stoch_chem}

\begin{thebibliography}{10}

\bibitem{AlfonsiCancesTuriniciVenturaHuisinga2005ESIAM}
\bl{Aur\'{e}lien Alfonsi, Eric Canc\`{e}s, Gabriel Turinici, Barbara Di Ventura, and Wilhelm Huisinga. 
\newblock Adaptive simulation of hybrid stochastic and deterministic models for biochemical systems.
\newblock In \emph{ESAIM: Proceedings}, vol.~14, pp.~1-13. EDP Sciences, 2005.}

\bibitem{Anderson2007JChemPhys}
David~F Anderson.
\newblock A modified next reaction method for simulating chemical systems with
  time dependent propensities and delays.
\newblock {\em J Chem Phys}, 127(21):214107, Dec 2007.

\bibitem{AndCFD2012}
David~F. Anderson.
\newblock An efficient finite difference method for parameter sensitivities of
  continuous time {M}arkov chains.
\newblock {\em SIAM Journal on Numerical Analysis}, 50(5):2237 -- 2258, 2012.

\bibitem{AndersonGangulyKurtz2011AnnApplPr}
David~F. Anderson, Arnab Ganguly, and Thomas~G. Kurtz.
\newblock Error analysis of tau-leap simulation methods.
\newblock {\em Ann. Appl. Probab.}, 21(6):2226--2262, December 2011.

\bibitem{AndHigham2012}
David~F. Anderson and Desmond~J. Higham.
\newblock Multi-level {M}onte {C}arlo for continuous time {M}arkov chains, with
  applications in biochemical kinetics.
\newblock {\em SIAM: Multiscale Modeling and Simulation}, 10(1):146 -- 179,
  2012.
  
  \bibitem{AHS2014}
  \bl{David~F. Anderson, Desmond~J. Higham, and Yu Sun.
  \newblock  Complexity Analysis of Multilevel {M}onte {C}arlo Tau-Leaping.
  \newblock Submitted, 2014.}
  
  \bibitem{AK2014}
  \bl{David~F. Anderson and Masanori Koyama.
  \newblock  An asymptotic relationship between coupling methods for stochastically modeled population processes.
  \newblock Accepted for publication to {\em IMA Journal of Numerical Analysis}, 2014.
  }

  
\bibitem{AndersonKurtz2010Chapter}
David~F. Anderson and Thomas~G. Kurtz.
\newblock {\em Design and Analysis of Biomolecular Circuits}, chapter 1.
  Continuous Time Markov Chain Models for Chemical Reaction Networks.
\newblock Springer, 2011.

\bibitem{AndSkubak}
David~F. Anderson and Elizabeth~Skubak Wolf.
\newblock A finite difference method for estimating second order parameter
  sensitivities of discrete stochastic chemical reaction networks.
\newblock {\em J. Chem. Phys.}, 137(22):224112, 2012.

\bibitem{Ball06}
Karen Ball, Thomas~G. Kurtz, Lea Popovic, and Greg Rempala.
\newblock Asymptotic analysis of multiscale approximations to reaction
  networks.
\newblock {\em Ann. Appl. Prob.}, 16(4):1925--1961, 2006.

\bibitem{Bressloff20009SIAM}
\bl{Paul C.~Bressloff. 
\newblock Stochastic neural field theory and the system-size expansion. 
\newblock \emph{SIAM Journal on Applied Mathematics} 70(5):1488-1521 (2009).}

\bibitem{BressloffNewby2013SIADS}
\bl{Paul C.~Bressloff and Jay M. Newby. 
\newblock Metastability in a stochastic neural network modeled as a velocity jump Markov process.
\newblock \emph{SIAM Journal on Applied Dynamical Systems} 12(3):1394-1435  (2013).}

\bibitem{BressloffNewby2014PhysBiol}
\bl{Paul C.~Bressloff and Jay M. Newby. 
\newblock Stochastic hybrid model of spontaneous dendritic NMDA spikes.
\newblock \emph{Physical biology} 11(1):016006 (2014).}

\bibitem{BressloffNewby2014PRE}
\bl{Paul C.~Bressloff and Jay M. Newby. 
\newblock Path integrals and large deviations in stochastic hybrid systems.
\newblock \emph{Physical Review E} 89(4):042701 (2014).}

\bibitem{BR:2011}
\bl{Evelyn Buckwar and Martin G. Riedler.
\newblock An exact stochastic hybrid model of excitable membranes including spatio-temporal evolution.
\newblock {\em J. Math. Biol.}, 63:1051--1093, 2011.}

\bibitem{BuiceCowanChow2010NeCo}
\bl{Michael A.~Buice, Jack D.~Cowan, and Carson C.~Chow. 
\newblock Systematic fluctuation expansion for neural network activity equations. \newblock \emph{Neural Computation} 22(2): 377-426 (2010).}

\bibitem{CaoGillespiePetzold2006JChemPhys}
Yang Cao, Daniel~T Gillespie, and Linda~R Petzold.
\newblock Efficient step size selection for the tau-leaping simulation method.
\newblock {\em J Chem Phys}, 124(4):044109, Jan 2006.

\bibitem{ClayDeFelice1983BiophysJ}
J~R Clay and L~J DeFelice.
\newblock Relationship between membrane excitability and single channel
  open-close kinetics.
\newblock {\em Biophys J}, 42(2):151--7, May 1983.

\bibitem{Colquhoun+Hawkes:1983chapter}
D.~Colquhoun and A.~G. Hawkes.
\newblock {\em Single-Channel Recording}, chapter The Principles of the
  Stochastic Interpretation of Ion-Channel Mechanisms.
\newblock Plenum Press, New York, 1983.

\bibitem{Davis:1984}  
\bl{M.~H.~A.~Davis.  
\newblock  Piecewise-Deterministic Markov Processes: A General Class of Non-Diffusion Stochastic Models.
\newblock {\em Journal of the Royal Statistical Society. Series B}, 46(3):353--388, 1984.}

\bibitem{Dorval+White:2005:JNsci}
Alan~D. {Dorval, Jr.} and John~A. White.
\newblock Channel noise is essential for perithreshold oscillations in
  entorhinal stellate neurons.
\newblock {\em The Journal of Neuroscience}, 25(43):10025--10028, Oct. 26 2005.

\bibitem{EarnshawKeener2010SIADS}
Berton~A. Earnshaw and James~P. Keener.
\newblock Invariant manifolds of binomial-like nonautonomous master equations.
\newblock {\em Siam J. Applied Dynamical Systems}, 9(2):568--588, 3 June 2010.

\bibitem{ErmentroutTerman2010book}
G.~Bard Ermentrout and David~H. Terman.
\newblock {\em Foundations Of Mathematical Neuroscience}.
\newblock Springer, 2010.


\bibitem{EK:1986}
\bl{Stewart N.~Ethier and Thomas G.~Kurtz.
\newblock {Markov Processes: Characterization and convergence.}
\newblock {John Wiley \& Sons}, New York, 1986.}


\bibitem{FischSchwalgerLindnerHerzBenda2012JNSci}
Karin Fisch, Tilo Schwalger, Benjamin Lindner, Andreas V~M Herz, and Jan Benda.
\newblock Channel noise from both slow adaptation currents and fast currents is
  required to explain spike-response variability in a sensory neuron.
\newblock {\em J Neurosci}, 32(48):17332--44, Nov 2012.

\bibitem{FoxLu1994PRE}
Ronald F. Fox and Yan-nan Lu.
\newblock Emergent collective behavior in large numbers of globally coupled
  independently stochastic ion channels.
\newblock {\em Phys Rev E Stat Phys Plasmas Fluids Relat Interdiscip Topics},
  49(4):3421--3431, Apr 1994.


\bibitem{Giles2008}
  Mike B. Giles.
 \newblock Multilevel {M}onte {C}arlo path simulation.
 \newblock {\em Operations Research}, 56:607--617, 2008.
 
\bibitem{Gillespie1977}
Daniel T.~Gillespie.
\newblock Exact stochastic simulation of coupled chemical reactions.
\newblock {\em J. Phys. Chem.}, 81:2340--2361, 1977.

\bibitem{Gillespie2007AnnRevPhysChem}
Daniel T.~Gillespie.
\newblock Stochastic simulation of chemical kinetics.
\newblock {\em Annu. Rev. Phys. Chem.}, 58:35--55, 2007.

\bibitem{Glynn89}
Peter~W. Glynn.
\newblock A {GSMP} formalism for discrete event systems.
\newblock {\em Proc. of the IEEE}, 77(1):14--23, 1989.

\bibitem{GoldwynImennovFamulareShea-Brown2011PRE}
Joshua~H Goldwyn, Nikita~S Imennov, Michael Famulare, and Eric Shea-Brown.
\newblock Stochastic differential equation models for ion channel noise in
  hodgkin-huxley neurons.
\newblock {\em Phys Rev E Stat Nonlin Soft Matter Phys}, 83(4 Pt 1):041908, Apr
  2011.

\bibitem{GoldwynShea-Brown2011PLoSCB}
Joshua~H Goldwyn and Eric Shea-Brown.
\newblock {The what and where of adding channel noise to the Hodgkin-Huxley
  equations}.
\newblock {\em PLoS Comput Biol}, 7(11):e1002247, Nov 2011.

\bibitem{GroffDeRemigioSmith2009Chapter}
\bl{Jeffrey R.~Groff, Hilary DeRemigio, and Gregory D. Smith.
\newblock Markov chain models of ion channels and calcium release sites.
\newblock In \emph{Stochastic Methods in Neuroscience} (2009): 29-64.
}

  

\bibitem{Haas2002}
Peter~J. Haas.
\newblock {\em Stochastic Petri Nets: Modelling Stability, Simulation}.
\newblock Springer, New York, first edition, 2002.

\bibitem{jp:Hodgkin+Huxley:1952d}
A.~L. Hodgkin and A.~F. Huxley.
\newblock A quantitative description of membrane current and its application to
  conduction and excitation in nerve.
\newblock {\em J Physiol}, 117:500--544, 1952.

\bibitem{KurtzKang:2013}
\bl{Hye-Won Kang and Thomas G. Kurtz.
\newblock{Separation of time-scales and model reduction for stochastic reaction models.}
\newblock{\em Ann. Appl. Probab.}, 23:529--583, 2013.}

\bibitem{KeenerNewby2011PRE}
James~P Keener and Jay~M Newby.
\newblock Perturbation analysis of spontaneous action potential initiation by
  stochastic ion channels.
\newblock {\em Phys Rev E Stat Nonlin Soft Matter Phys}, 84(1-1):011918, Jul
  2011.

\bibitem{KisperskyTilman2008Scholarpedia}
T.~Kispersky and J.~A. White.
\newblock Stochastic models of ion channel gating.
\newblock {\em Scholarpedia}, 3(1):1327, 2008.


\bibitem{Kurtz80}
Thomas G. Kurtz.
\newblock Representations of Markov Processes as Multiparameter Time Changes.
\newblock {\em Ann. Prob.}, 8(4):682--715, 1980.


\bibitem{KurtzPop81}
Thomas G. Kurtz.  
\newblock  {\em Approximation of population processes}, {CBMS-NSF Reg. Conf. Series in Appl. Math.: 36}, SIAM, 1981.



\bibitem{LaingLord2010}
Carlo Laing and Gabriel~J. Lord, editors.
\newblock {\em Stochastic Methods in Neuroscience}.
\newblock Oxford University Press, 2010.

\bibitem{LeeOthmer2010JMB}
Chang Lee and Hans Othmer.
\newblock A multi-time-scale analysis of chemical reaction networks: I.
  deterministic systems.
\newblock {\em Journal of Mathematical Biology}, 60:387--450, 2010.
\newblock 10.1007/s00285-009-0269-4.

\bibitem{MilescuYamanishiPtakSmith2010JNsci}
Lorin~S Milescu, Tadashi Yamanishi, Krzysztof Ptak, and Jeffrey~C Smith.
\newblock Kinetic properties and functional dynamics of sodium channels during
  repetitive spiking in a slow pacemaker neuron.
\newblock {\em J Neurosci}, 30(36):12113--27, Sep 2010.

\bibitem{MinoRubinsteinWhite2002AnnBiomedEng}
Hiroyuki Mino, Jay~T Rubinstein, and John~A White.
\newblock Comparison of algorithms for the simulation of action potentials with
  stochastic sodium channels.
\newblock {\em Ann Biomed Eng}, 30(4):578--87, Apr 2002.

\bibitem{MorrisLecar1981BiophysJ}
C.~Morris and H.~Lecar.
\newblock Voltage oscillations in the barnacle giant muscle fiber.
\newblock {\em Biophysical Journal}, 35(1):193 -- 213, 1981.

\bibitem{NewbyBressloffKeener2013PRL}
Jay~M Newby, Paul~C Bressloff, and James~P Keener.
\newblock Breakdown of fast-slow analysis in an excitable system with channel
  noise.
\newblock {\em Phys Rev Lett}, 111(12):128101, Sep 2013.

\bibitem{PTW:2010}
\bl{Khashayar Pakdaman, Mich\`ele Thieullen, and Gilles Wainrib.
\newblock Fluid limit theorems for stochastic hybrid systems with application to neuron models.
\newblock {\em Adv. Appl. Prob.}, 42(3):761--794, Sept. 2010. }

\bibitem{PTW:2012}
\bl{Khashayar Pakdaman, Mich\`ele Thieullen, and Gilles Wainrib.
\newblock Asymptotic expansion and central limit theorem for multiscale piecewise-deterministic Markov processes.
\newblock {\em Stoch. Proc. Appl.}, 122:2292--2318, 2012. }


\bibitem{Khammash2010}
Muruhan Rathinam, Patrick~W. Sheppard, and Mustafa Khammash.
\newblock Efficient computation of parameter sensitivities of discrete
  stochastic chemical reaction networks.
\newblock {\em Journal of Chemical Physics}, 132:034103, 2010.

\bibitem{Riedler2013}
Martin Riedler and Girolama Notarangelo.
\newblock Strong Error Analysis for the $\Theta$-Method for Stochastic Hybrid Systems.
\newblock {\em arXiv preprint arXiv:1310.0392}, 2013

\bibitem{RiedlerThieullenWainrib2012ElecJProb}
\bl{Martin G. Riedler, Mich\`{e}le Thieullen, and Gilles Wainrib.
\newblock Limit theorems for infinite-dimensional piecewise deterministic Markov processes. Applications to stochastic excitable membrane models.
\newblock {\em Electron.~J.~Probab.}, 17(55),1-48 (2012).}

\bibitem{Rinzel+Ermentrout:1989}
J.~Rinzel and G.B. Ermentrout.
\newblock Analysis of neural excitability and oscillations.
\newblock In C.~Koch and I.~Segev, editors, {\em Methods in Neuronal Modeling}.
  MIT Press, second edition, 1989.

\bibitem{Rubin+Terman:2002}
D.~Terman J.~Rubin.
\newblock Geometric singular pertubation analysis of neuronal dynamics.
\newblock In B.~Fiedler, editor, {\em Handbook of Dynamical Systems, vol. 2:
  Towards Applications}, pages 93--146. Elsevier, 2002.

\bibitem{SchmandtGalan2012PRE}
Nicolaus~T Schmandt and Roberto~F Gal\'{a}n.
\newblock Stochastic-shielding approximation of {Markov} chains and its
  application to efficiently simulate random ion-channel gating.
\newblock {\em Phys Rev Lett}, 109(11):118101, Sep 2012.

\bibitem{SchmidtThomas2014JMN}
Deena~R.~Schmidt and Peter~J.~Thomas.
\newblock Measuring edge importance: a quantitative analysis of the stochastic shielding approximation for random processes on graphs.
\newblock {\em The Journal of Mathematical Neuroscience}, 4(1):6, 2014.

\bibitem{SchwalgerFischBendaLindner2010PLosCB}
Tilo Schwalger, Karin Fisch, Jan Benda, and Benjamin Lindner.
\newblock How noisy adaptation of neurons shapes interspike interval histograms
  and correlations.
\newblock {\em PLoS Comput Biol}, 6(12):e1001026, 2010.

\bibitem{ShingaiQuandt1986BrainRes}
R~Shingai and FN~Quandt.
\newblock Single inward rectifier channels in horizontal cells.
\newblock {\em Brain Research}, 369(1-2):65--74, Mar 26 1986.

\bibitem{SkaugenWalloe1979ActaPhysiolScand}
E~Skaugen and L~Wall{\o}e.
\newblock Firing behaviour in a stochastic nerve membrane model based upon the
  {{Hodgkin-Huxley}} equations.
\newblock {\em Acta Physiol Scand}, 107(4):343--63, Dec 1979.

\bibitem{SmithKeizer2002BookChapter}
Gregory D.~Smith and Joel Keizer.
\newblock Modeling the stochastic gating of ion channels.
\newblock In \emph{Computational Cell Biology}, pp. 285-319. Springer New York, 2002.

\bibitem{rishi2013}
Rishi Srivastava, David F. Anderson, and James B. Rawlings. 
\newblock Comparison of finite difference based methods to obtain sensitivities of stochastic chemical kinetic models.
\newblock {\em Journal of Chemical Physics}, 138(7):074110, 2013.


\bibitem{StrassbergDeFelice1993NeCo}
Adam~F. Strassberg and Louis~J. {DeFelice}.
\newblock Limitations of the {Hodgkin-Huxley} formalism: Effects of single
  channel kinetics on transmebrane voltage dynamics.
\newblock {\em Neural Computation}, 5:843--855, 1993.



\bibitem{WTP:2012}
\bl{Gilles Wainrib,  Mich\`ele Thieullen, and Khashayar Pakdaman.
\newblock Reduction of stochastic conductance-based neuron models with time-scales separation.
\newblock {\em J. Comput. Neurosci.}, 32:327--346, 2012. }




\bibitem{White+Chow+Ritt+Soto-Trevino+Kopell:1998:JCN}
J.A. White, C.C. Chow, J.~Ritt, C.~Soto-Trevino, and N.~Kopell.
\newblock Synchronization and oscillatory dynamics in heterogeneous, mutually
  inhibited neurons.
\newblock {\em Journal of Computational Neuroscience}, 5:5--16, 1998.

\bibitem{White+Kay+Rubinstein:2000:TrendsNsci}
J.A. White, J.T. Rubinstein, and A.R. Kay.
\newblock Channel noise in neurons.
\newblock {\em Trends Neurosci.}, 23:131--137, 2000.

\bibitem{Wilkinson2011}
Darren~J. Wilkinson.
\newblock {\em Stochastic Modelling for Systems Biology}.
\newblock Chapman \& Hall/CRC, Nov 2011.

\end{thebibliography}

\end{document}